\RequirePackage{fix-cm}
\documentclass{article}                      
\usepackage{amsmath}\usepackage{amsmath,amsfonts,amssymb,amsthm,url,array}
\usepackage{mathbbol}
\usepackage{abstract}
\usepackage{hyperref}
\usepackage[utf8]{inputenc}
\usepackage[T1]{fontenc}

\RequirePackage[numbers,round]{natbib}
\usepackage{setspace}
\usepackage{mathptmx}
\setcitestyle{square}
\usepackage{titlesec}
\titleformat{\paragraph}
{\normalfont\normalsize\bfseries}{\theparagraph}{1em}{}
\titlespacing*{\paragraph}
{0pt}{3.25ex plus 1ex minus .2ex}{1.5ex plus .2ex}
\usepackage{optidef}
\usepackage{graphicx}
\usepackage{caption}
\usepackage{subcaption}
\usepackage[nomarkers]{endfloat}

\usepackage{caption}
\usepackage{blindtext}
\usepackage{fancyhdr}
 
\begin{document}

\title{Discrete Time Portfolio Optimization managing Value at Risk \\ under heavy tail return distribution}

\author{
   Subhojit Biswas\\
    \textit{Indian Statistical Institute, Kolkata}\\
    \textit{subhojit1016kgp@gmail.com}\\
    \\
    Diganta Mukherjee\thanks{We thank Mrinal K. Ghosh for helpful comments and suggestions which has helped improving the exposition considerably. The usual caveat applies.}\\
    \textit{Sampling and Official Statistics Unit}\\
    \textit{Indian Statistical Institute, Kolkata}\\
    \textit{digantam@hotmail.com}}

\maketitle

\setcounter{page}{1}

\begin{abstract}
\noindent We consider an investor, whose portfolio consists of a single risky asset and a risk free asset, who wants to maximize his expected utility of the portfolio subject to the Value at Risk assuming a heavy tail distribution of the stock price return. We use Markov Decision Process and dynamic programming principle to get the optimal strategies and the value function which maximize the expected utility for parametric as well as non parametric distributions. Due to lack of explicit solution in the non-parametric case, we use numerical integration for optimization. 
\section*{Highlights}
\begin{itemize}
   \item Used Markov Decision Process and dynamic programming to get recursive relation for optimal strategy of known distribution
   \item Once it approaches the terminal position we will build up more on the risky asset or will liquidate the risky asset
   \item In the non-parametric case, we used numerical integration to find the recursive relation for optimal strategy
   \item In this case the value function we obtained has wide range of fluctuations and the same is also true for the portfolio wealth
\end{itemize}
\end{abstract}

\noindent {\bf Keywords: Portfolio Optimization, Markov Decision Process, Parametric distribution,  Non-parametric distribution}

\noindent {\bf AMS Classification: 91G10, 91G80}

\noindent {\bf Biographical Notes:} Mr. Subhojit Biswas, completed his M.Tech in Quality, Reliability and Operations Research at Indian Statistical Institute. Previously, he has completed his B.Tech in Instrumentation Engineering from Indian Institute of Technology, Kharagpur. Research interests are Mathematical Finance, Optimization and Applied Mathematics. Worked at Barclays for 3 years before coming for his masters. Currently working at JPMorgan Chase.\newline
\noindent Dr. Diganta Mukherjee, Professor at Sampling and Official Statistics Unit of Indian Statistical Institute, Kolkata completed his BStat, MStat, PhD (Economics) all from Indian Statistical Institute, Kolkata. Research interests are Welfare \& Development Economics and Finance. Ex-faculty, Jawaharlal Nehru University, Essex University, ICFAI Business School. Has over 80 publications in national and international journals and authored four books. Has been involved in projects with large corporate houses and various ministries of GoI and WBGovt. Acting as technical advisor to MCX, RBI, SEBI, NSSO, NAD (CSO).

\section{Introduction}

\subsection{Background and Motivation}

Risk management occurs everywhere in the financial world. It is done when an investor buys low-risk government bonds over riskier corporate bonds, bank performing a credit check on an individual before issuing a personal line of credit, stockbrokers buying assets like options $\&$ futures in their portfolio and money managers using strategies like portfolio and investment diversification to mitigate or effectively manage risk. Inadequate risk management can result in severe consequences such as the sub prime mortgage meltdown in 2007 that helped trigger the Great Recession stemming from poor risk-management decisions. In the financial world the performance measure of the portfolio associated with risk and portfolio management is actually risk management. A common definition of investment risk is a deviation from an expected outcome, which we can benchmark with the market parameters. The deviation can be positive or negative. How Do Investors Measure Risk?
Investors use a variety of tactics to ascertain risk. One of the most commonly used risk metrics, Value at Risk (VaR), is a statistical measure of the riskiness of financial entities or portfolios of assets. It is defined as the maximum dollar amount expected to be lost over a given time horizon, at a pre-defined confidence level. There are also other risk measure metrics used in the market such as Sharpe's Ratio or Expected Shortfall (ES). As mentioned, our main focus in this paper will be Value at Risk (VaR).

\noindent Consider an investor who is interested to optimize the portfolio return while managing VaR. It is well documented that returns on financial assets often exhibit heavy tails, so much so that even the first moment may not exist for the return distribution (see \citeauthor{CLM}, \citeyear{CLM}). 
In this paper we consider investment strategies that maximize the median return of the portfolio such that VaR is controlled for a given lower quantile level.
We incorporate the effect of transaction cost and do the analysis for known and  unknown distribution of returns.

\subsection{Literature Review}

Interest rate risk immunization is one of the key concerns for fixed income portfolio management. In recent years, risk measures (e.g. value‐at‐risk and conditional value‐at‐risk)is used as tools for the formation of an optimum investment portfolio has received much attention. The article by \citeauthor{Ref5}(\citeyear{Ref5}) aims to discuss this issue. The paper by \citeauthor{Ref6} (\citeyear{Ref6}) empirically test the performance of different models in measuring VaR and ES in the presence of heavy tails in returns using historical data. Daily returns are modelled with empirical (or historical), Gaussian or Generalized Pareto (peak over threshold (POT) technique of extreme value theory (EVT)) distributions. Assessing financial risk and portfolio optimization using a multivariate market model with returns assumed to follow a multivariate normal tempered stable distribution (i.e. a mixture of the multivariate normal distribution and the tempered stable subordinator) can be seen in \citeauthor{Ref7}(\citeyear{Ref7}). Several authors have considered the optimal portfolio problem under drawdown constraint. The first to comprehensively study this problem over infinite time horizon in a market setting with single risky asset modelled as a geometric Brownian motion with constant volatility (log normal model) was \citeauthor{Ref8}(\citeyear{Ref8}). Dynamic programming was used to solve the maximization problem of long term growth rate of the expected utility of wealth. \citeauthor{Ref9}(\citeyear{Ref9}) streamlined the analysis of \citeauthor{Ref8}(\citeyear{Ref8}) and extended the results to the case when there are multiple risky assets. The pioneering paper by  \citeauthor{Ref10}(\citeyear{Ref10}) gives an idea of portfolio selection by stochastic dynamic programming. The paper by \citeauthor{Ref12}(\citeyear{Ref12}) gives an idea of using a non parametric estimator for the State Price Densities implicit in option prices. Mean-VaR  portfolio optimization is studied in the paper by \citeauthor{Ref14}(\citeyear{Ref14}) in a non-parametric setting. In this paper the authors have investigated the portfolio optimization problem with six practical constraints widely used in real life trading scenarios. The paper by \citeauthor{Ref15}(\citeyear{Ref15}) gives an idea of VaR constrained Markowitz style portfolio selection problem, here the distribution of the returns of considered assets are given in the form of finitely many scenarios. Numerical analysis is used to solve the problem. When the utility function is quadratic then how to use the Mean-VaR portfolio optimization is described in the paper by \citeauthor{Ref13}(\citeyear{Ref13}). In the paper by \citeauthor{Ref16}(\citeyear{Ref16}), the authors have devised an algorithm to solve conditional variance problem by Markov Decision Process. The book by \citeauthor{Ref17}(\citeyear{Ref17}) gives us the idea of how to use stochastic recursive algorithm for optimization problems. The paper by \citeauthor{Ref19}(\citeyear{Ref19}) considers statistical Markov Decision Processes where the decision maker is risk averse against model ambiguity. The model ambiguity is given by an unknown parameter which influences the transition law and the cost functions. Risk aversion is either measured by the entropy risk measure or by the average Value at Risk. Minimax optimization is used to solve this problem. The paper by \citeauthor{Ref20}(\citeyear{Ref20}) uses the approximate dynamic programming to set up a Markov decision model for the multi-time segment portfolio with transaction cost. The paper by \citeauthor{Ref21}(\citeyear{Ref21}) analyses the contract between an entrepreneur and an investor, using a non-zero sum game in which the entrepreneur is interested in company survival and the investor in maximizing expected net present value. This paper looks into a different setup for a finance company. The paper by \citeauthor{Ref22}(\citeyear{Ref22}) analyses the data of the CSI 300 Index in the past five years, and uses Monte Carlo simulation and historical simulation to calculate the VaR of the five-year index and test its validity. It combines the result with China’s market economy and puts forward some suggestions on financial risk management in China’s financial market. The main contribution of the paper by \citeauthor{Ref23}(\citeyear{Ref23}) is to analyze the application of multi-state Markov models to evaluate credit card risk by investigating the characteristics of different state transitions in client-institution relationships over time, thereby generating score models for various purposes. It gives a different direction of application of the Markov Decision Process in the financial market. The paper by \citeauthor{Ref24}(\citeyear{Ref24}) considers the variance optimization problem of average reward in continuous-time Markov decision process (MDP). It assumes that the state space is countable and the action space is a Borel measurable space. The main purpose of the paper is to find the policy with the minimal variance in the deterministic stationary policy space. The paper by \citeauthor{Ref25}(\citeyear{Ref25}) studies a portfolio optimization problem combining a continuous-time jump market and a defaultable security and provides a numerical solutions through the conversion into a Markov decision process. The paper also analyse allocation strategies under several families of utility functions and compares it with previously obtained results.

\subsection{Our Contribution}

In this article we consider an investor who is worried about when to build up on stocks or liquidate the stock when dealing with heavy tail distribution of the return of the stock prices, while controlling Value at Risk (VaR). The investor's portfolio has one risky asset and a risk free asset. We consider this problem in discrete time and then apply the general Markov Decision Problem formulation. Markov decision process theory and algorithms for finite horizon problems primarily concerns with determining a policy with the largest expected total reward. We try to determine the policies for two scenarios where the distribution is known and when the distribution is not known. Pareto, Weibull and Inverse Gaussian distributions are used for the parametric scenario and for the non-parametric case we have used the kernel density estimator to fit our data and carry out the numerical integration.

\subsection{Organization of the paper}
In section 2 we outline our approach for portfolio optimization where we discretized the wealth equation and applied the general Markov Decision Problem formulation. So it has become a dynamic programming problem. We tried to find the optimal strategy considering two scenarios where the distribution is known (section 3) and when the distribution is not known (section 4). The candidate known distributions are Pareto, Weibull and Inverse Gaussian. We also make a comparative goodness of fit discussion among these three and two other distributions commonly used in the literature (mixture normal and variance gamma, as discussed above) in a separate subsection 3.4 of section 3.
For non-parametric case, kernel density estimator is used to fit our data and carry out the numerical integration. We interpret the results under different circumstances, with and without transaction cost. Section 5 concludes the article. 

\section{Methodology}

We work with one risky asset $S_t$ and a risk free bank account with return r. Again at a certain point of time we are investing $\pi_t$ of the total investible wealth $M$ in the risky asset and the rest in the bank account (so $\pi_t \in [0, M]$). We can express the equation of wealth as 
\begin{equation}\label{eq25}
    L_t = \pi_{t} S_t + (M - \pi_{t}) r.
\end{equation}

\noindent The heavy tail distribution doesn't allow us to use the properties of Brownian motion and the Hamilton-Jacobi-Bellman (HJB) equation to solve the stochastic differential equations to obtain the optimal strategy and the value function in continuous time. To obtain a solution in such a situation, we convert this problem into a discrete domain and then apply the general Markov Decision Problem (MDP) formulation. Since we have a fixed time horizon, it is treated as a Finite Horizon Markov Decision Process. Markov decision process theory and algorithms for finite horizon problems is primarily concerned with determining a policy $\pi^{\ast}$ with the largest expected total reward. We seek a policy $\pi^{\ast}$ for which (see \citeauthor{Ref4},\citeyear{Ref4})
\begin{equation}
    v_N^{\pi^\ast}(s) \ge v_N^{\pi}(s), \nonumber
\end{equation}
where $s\in S$ is the state of nature. Finite-horizon policy evaluation algorithm is used  to compute $v_N^{\pi}$. To start with let
$u_t^{\pi}$ denote the total expected reward obtained by using policy $\pi$ at decision epochs $t, t+1, t+2,...,N-1$. If the history at decision epochs is $h_t$ then $u_t^{\pi}$ is defined as

\begin{equation}
    u_t^{\pi}(h_t) = \mathop{\mathbb{E}}_{h_t}^{\pi}\big(\sum_{n=t}^{N-1}W_n + W_N \big ), \nonumber
\end{equation}
where $W_n$ denotes the reward. The difference between $u_t^{\pi}$ and $v_N^{\pi}$ is that $v_N^{\pi}$ includes rewards over the entire future, whereas $u_t^{\pi}$ only incorporates rewards from decision epoch t onward. Here we can use the algorithm from (\citeauthor{Ref4},\citeyear{Ref4}) as below. Consider the formulation
\begin{equation}
    u_{t-1}^{\pi}(h_{t-1}) = W_{t-1}(s_{t-1},d_{t-1}(h_{t-1})) + \sum_{j\in S}p_{t-1}(j|s_{t-1},d_{t-1}(h_{t-1}))u_{t-1}^{\pi}(h_{t-1},d_{t-1}(h_{t-1}),j).\nonumber
\end{equation}
The expected value of policy $\pi$ over periods $t, t+1,....$, when the history of epoch t is $h_{t-1}$, equals the immediate reward received by selecting the action $d_t(h_{t-1})$ plus the expected reward over the remaining time periods. The second term contains the product of the probability of being in state j at decision epoch $t+1$ if action $d_t(h_{t-1})$ is used and the median total reward obtained using the policy $\pi$ over periods t, t+1,..., N when the history at epoch t is $h_t = (h_{t-1}, d_{t-1}(h_{t-1}), j)$. Summing over all possible j give the desired expectation expressed in terms of  
\begin{equation}
    u_{t-1}^{\pi}(h_{t-1}) = W_{t-1}(s_{t-1},d_{t-1}(h_{t-1})) + \mathop{\mathbb{E}}^{\pi}_{h_{t-1}}\{u_{t}^{\pi}(h_{t-1}, d_{t-1}(h_{t-1}), X_t)\} \nonumber
\end{equation}
where $X_t$ determines the state at time t. The solution of these equations correspond to optimal value functions and they also provide a basis for determining the optimal policies. 
\begin{equation}
    u^{*}_{t-1}(h_{t-1}) = \sup_{\pi\in \Pi^{HR}}  \Big\{u_{t-1}^{\pi}(h_{t-1})\Big\} , \nonumber
\end{equation}
where HR is the randomized history dependent policy. Expanding the R.H.S of the above equation, we get
\begin{equation}\label{eq26}
    u^{\pi^{*}}_{t-1}(h_{t-1}) = \sup_{a\in A_{s_t}}  \Big\{W_{t-1}(s_{t-1}, a) + \sum_{j\in S}p_{t-1}(j|s_{t-1}, a)u_{t}^{\pi}(h_{t-1}, a, j)\Big\} . 
\end{equation}
In our scenario we have policy $\pi$ which is a mapping between the state action couple. The action $A_{s_t}$ has two alternatives - whether to put money in the stock or withdraw money from the stocks - therefore making it finite and at the same time the state is different across time periods, so we can replace the supremum by maximum and our optimization problem takes the following form
\begin{maxi}|l|
	  {{\pi:s->a},s\in (t, t+1,...T-1)}{\Big\{u_{t-1}^{\pi}(h_{t-1})\Big\}}{}{}
	  \addConstraint{P (L_{t-1} > Q_{0.05} | L_{t} > Q_{0.05}) }{\ge 0.95 ,}{}
	  \nonumber
\end{maxi}
which can be further rewritten as,
\begin{maxi}|l|
	  {{\pi:s->a},s\in (t, t+1,...T-1)}{\Big\{W_{t-1}(s_{t-1}, a) + \sum_{j\in s}p_{t-1}(j|s_{t-1}, a)u_{t}^{\pi}(h_{t-1}, a, j)\Big\}}{}{}
	  \addConstraint{P (L_{t-1} > Q_{0.05} | L_{t} > Q_{0.05}) }{\ge 0.95 .}{}
	  \nonumber
\end{maxi}

\noindent We have considered a quantile of 0.05 for VaR in this paper throughout. The constraint equation will help to identify the recursive relation between each optimal strategy $\pi$ at each time instant and this optimal strategy will help to maximize the value function. Basically we are going to use \eqref{eq26} for the value function as mentioned before and the reward is the change in the wealth from one time period to another time period.
\begin{equation}
    W_{t-1} = L_{t-1} - L_t \nonumber
\end{equation}
and for each $\pi_{t-1}$ obtained for a period the maximum value of $W_{t-1}$ is calculated which in turn helps to maximizes $u_{t-1}^{\pi}$.

\begin{equation}
    u_{t-1}^{\pi} = (L_{t-1} - L_t) + \sum p(t-1| t) u_{t}^{\pi} \nonumber
\end{equation}
\begin{equation}\label{eq28}
    u_{t-1}^{\pi} = (L_{t-1} - L_t) + \mathop{\mathbb{E}} (u_{t}^{\pi}) 
\end{equation}

\section{Examples and numerical results for known distribution of return}

For the numerical illustration, data used are daily closing price of the stock “Entergy Corporation” in the time range 31st August, 2009 till 30th August, 2013 (\citeauthor{Quantopian}, \citeyear{Quantopian}). The return is calculated for this data using the following expression
\begin{equation}
    return = \frac{(Price_{t} - Price_{t-1})}{Price_{t-1}} \nonumber
\end{equation}.

\noindent We chose three benchmark distributions for our parametric illustration. The first candidate is the Pareto distribution which is a the canonical example of power-law that is used in description of social, scientific, geophysical, actuarial and other observable phenomena. The focus being to describe the distribution of wealth in a society, fitting the trend that a large portion of wealth is held by a small fraction of the population. We next use the Weibull distribution which is one of the three canonical extreme value distributions (Frechet and Gumbell being the other two. For details refer to \citeauthor{EKM}, \citeyear{EKM}). The last candidate is the Inverse Gaussian which is a common candidate for modelling processes with jumps. As financial asset returns are nowadays thought of as having discontinuities (in the form of discrete jumps), the use of Levy process based distributions have gained currency. For details see \citeauthor{Kyprianou}, \citeyear{Kyprianou}.  Inverse Gaussian is a popular candidate of this class of distributions.  

\subsection{Pareto Distribution}
For a random variable with a Pareto (Type I) distribution, the probability that X is greater than some number x is given by
\begin{equation}
 {\overline {F}}(x)=\Pr(X>x)={\begin{cases}\left({\frac {\lambda}{x}}\right)^{\alpha }&x\geq \lambda,\\1&x<\lambda,\end{cases}}\nonumber
\end{equation}
\noindent The Pareto Type I distribution is characterized by a scale parameter $\lambda$ and a shape parameter $\alpha$, which is known as the tail index. When this distribution is used to model the distribution of wealth, then the parameter $\alpha$ is called the Pareto index.
We are going to fit a Pareto distribution to the data-set and estimate the scale and shape parameters of the Pareto distribution using the Maximum Likelihood Estimator (MLE). 
The likelihood function for the Pareto distribution parameters $\alpha$ and $\lambda$, given an independent sample $x = (x_1, x_2, ..., x_n)$, is
\begin{equation}
 L(\alpha ,\lambda)=\prod _{i=1}^{n}\alpha {\frac {\lambda^{\alpha }}{x_{i}^{\alpha +1}}}=\alpha ^{n}\lambda^{n\alpha }\prod _{i=1}^{n}{\frac {1}{x_{i}^{\alpha +1}}}\nonumber
 \end{equation}
 \newline
Therefore, the logarithmic likelihood function is
\begin{equation}
\ell (\alpha ,\lambda)=n\ln \alpha +n\alpha \ln \lambda-(\alpha +1)\sum _{i=1}^{n}\ln x_{i} \nonumber
\end{equation}
It can be seen that $\ell (\alpha ,x_{\mathrm {m} })$ is monotonically increasing with $\lambda$, that is, the greater the value of $\lambda$, the greater the value of the likelihood function. Hence, since $x \geq \lambda$, we conclude that

\begin{equation}
{\widehat \lambda}=\min _{i}{x_{i}}\nonumber
\end{equation}
To find the estimator for $\alpha$ we compute the corresponding partial derivative and determine where it is zero,
\begin{equation}
{\frac {\partial \ell }{\partial \alpha }}={\frac {n}{\alpha }}+n\ln \lambda-\sum _{i=1}^{n}\ln x_{i} = 0 \nonumber
\end{equation}
Thus the maximum likelihood estimator for $\alpha$ is:
\begin{equation}
{\widehat {\alpha }}={\frac {n}{\sum _{i}\ln(x_{i}/{\widehat \lambda})}}\nonumber
\end{equation}
The expected standard error of the parameters are
\begin{equation}
\sigma ={\frac {\widehat {\alpha }}{\sqrt {n}}} \nonumber 
\end{equation}
Now calculating the parameters along with the standard error after fitting the returns in Pareto distribution, we calculate the quantile $Q_{0.05}$ for the distribution. The constraint is,
\begin{equation}\label{eq29}
   \frac{ P (L_{t-1} > Q_{0.05},\\ L_{t} > Q_{0.05})}{P (L_{t} > Q_{0.05})} \ge 0.95 
\end{equation}
The scale and shape parameters calibrated from data are as follows (refer Table \ref{pareto}).\newline
\centerline{\textbf{Table \ref{pareto} should be here}} 

\noindent For judging the out of sample performance of our calibration exercise, once the parameters are estimated using 700 return data points, we check whether the next 50 data points are coming from the same distribution. For this we used the one-sample Kolmogorov-Smirnov statistic to test whether the sample comes from the Pareto distribution. The Kolmogorov–Smirnov statistic quantifies a distance between the empirical distribution function of the sample and the reference cumulative distribution function. Assuming the null hypothesis holds and the data follows a Pareto distribution, we use the following test statistic:\newline
\noindent The empirical distribution function $F_n$ for n independent and identically distributed (i.i.d.) ordered observations $X_i$ is defined as
\begin{equation}
F_{n}(x)={1 \over n}\sum _{i=1}^{n}I_{[-\infty ,x]}(X_{i})\nonumber
\end{equation}
where $I_{[-\infty ,x]}(X_{i})$ is the indicator function, equal to 1 if $ X_{i}\leq x$ and equal to 0 otherwise.\newline
\noindent The Kolmogorov–Smirnov statistic for a given cumulative distribution function F(x) is
\begin{equation}
D_{n}=\sup _{x}|F_{n}(x)-F(x)|. \nonumber
\end{equation}
We obtain the test statistic value $D_{50} = 0.16816$, which is less than the critical value at 5\% level $D_{50,0.05} = 0.19206$. Since $D_{50} < D_{50,0.05}$, we cannot reject the null hypothesis and conclude that there is no significant difference between distribution of the data and Pareto distribution (with $\lambda$ = 85.34364 and $\alpha$ = 10346.37374).\newline

\noindent The rate of interest (daily) is taken to be one of three alternative values. These are 0.00008 (equivalent to an annual rate of 2\%), 0.00014 (3.5\%) and 0.00024 (6\%).\newline
Putting the value of the quantile in equation \eqref{eq29} and putting the value of the wealth from \eqref{eq25} in to the same equation we get,

\begin{equation}
  \frac{ P (\pi_{t-1}S_{t-1} + (M-\pi_{t-1})r > 5\times 10^{-6}\lambda, \pi_{t}S_{t} + (M-\pi_{t})r > 5\times 10^{-6}\lambda)}{P (\pi_{t}S_{t} + (M-\pi_{t})r > 5\times 10^{-6}\lambda)} \ge 0.95 \nonumber 
\end{equation}
Taking M = 1 we get
\begin{equation}\
  \frac{ P \big(S_{t-1} > \frac{5\times 10^{-6}\lambda + r(\pi_{t-1}-1)}{\pi_{t-1}}, S_{t} > \frac{5\times 10^{-6}\lambda + r(\pi_{t}-1)}{\pi_{t}}\big)}{P \big(S_{t} > \frac{5\times 10^{-6}\lambda + r(\pi_{t}-1)}{\pi_{t}}\big)} \ge 0.95  \nonumber
\end{equation}
For simplicity let us denote the upper and lower limits by the following constants,

\begin{center}
\begin{tabular}{c c l l}
     $K_{t}$ &=& $\frac{5\times 10^{-6}\lambda + r(\pi_{t}-1)}{\pi_{t}}$ 
\end{tabular}
\end{center}
Then the constraint becomes
\begin{equation} \label{eq30}
    \frac{\int\limits_{K_{t-1}}^{K_{t}} \frac{\alpha\lambda^{\alpha}}{(x + \lambda)^{\alpha +1}} dx}{1-\int\limits_{0}^{K_t} \frac{\alpha\lambda^{\alpha}}{(x + \lambda)^{\alpha +1}} dx}\ge 0.95
\end{equation}
Solving \eqref{eq30} and putting the values of $\lambda$ and $r = 0.00014$ we get the recursive relation between the optimal strategies for t to t-1,

\begin{equation}\label{eq31}
  \pi_{t-1} \le \frac{\pi_t}{(19.1977 \pi_t + 1)}.  
\end{equation}

\begin{figure}
\centering
\begin{subfigure}{.6\textwidth}
  \centering
  \includegraphics[width=0.95\textwidth]{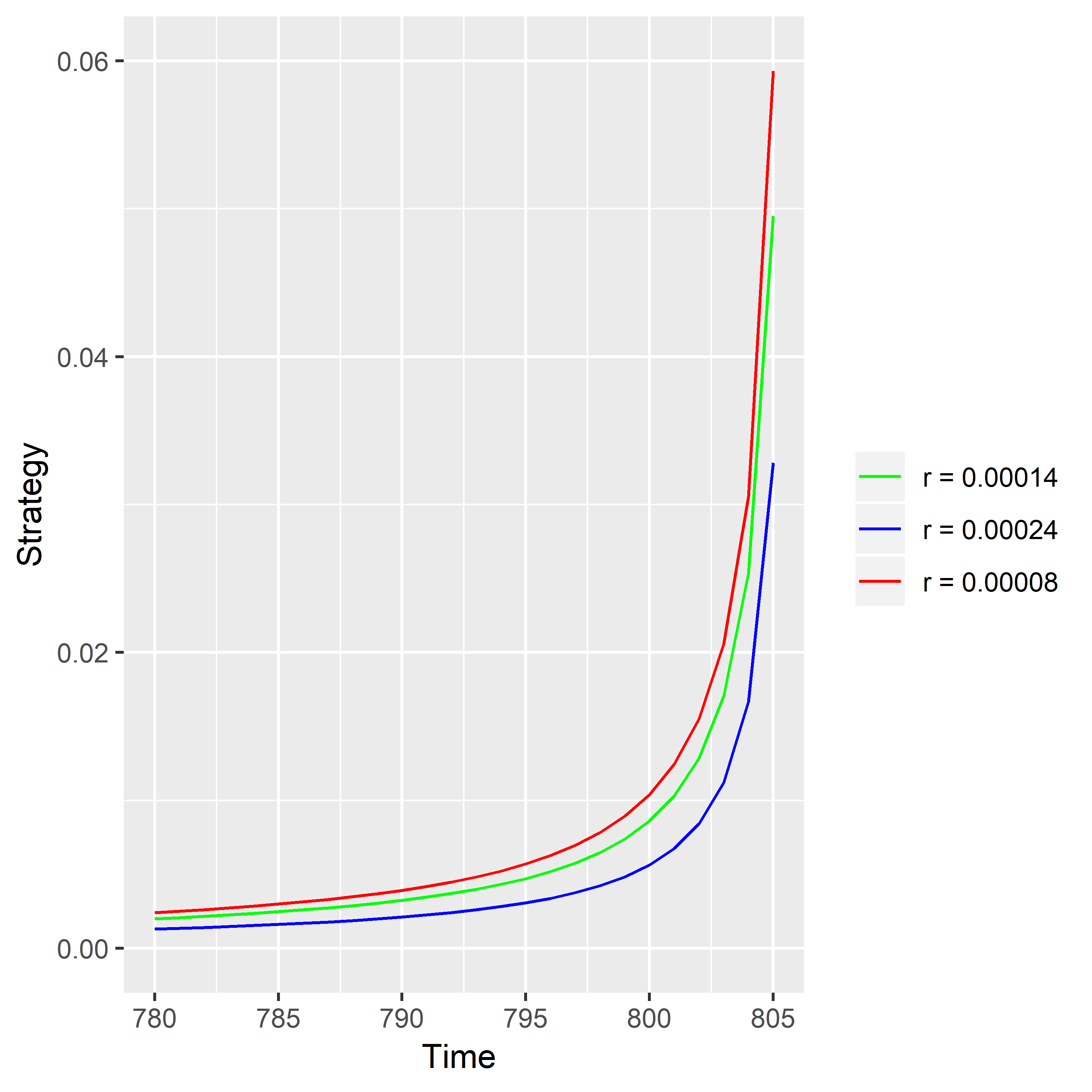}
\caption{When there is no transaction cost }
\label{fig:4}
\end{subfigure}%
\begin{subfigure}{.6\textwidth}
  \centering
  \includegraphics[width=0.95\textwidth]{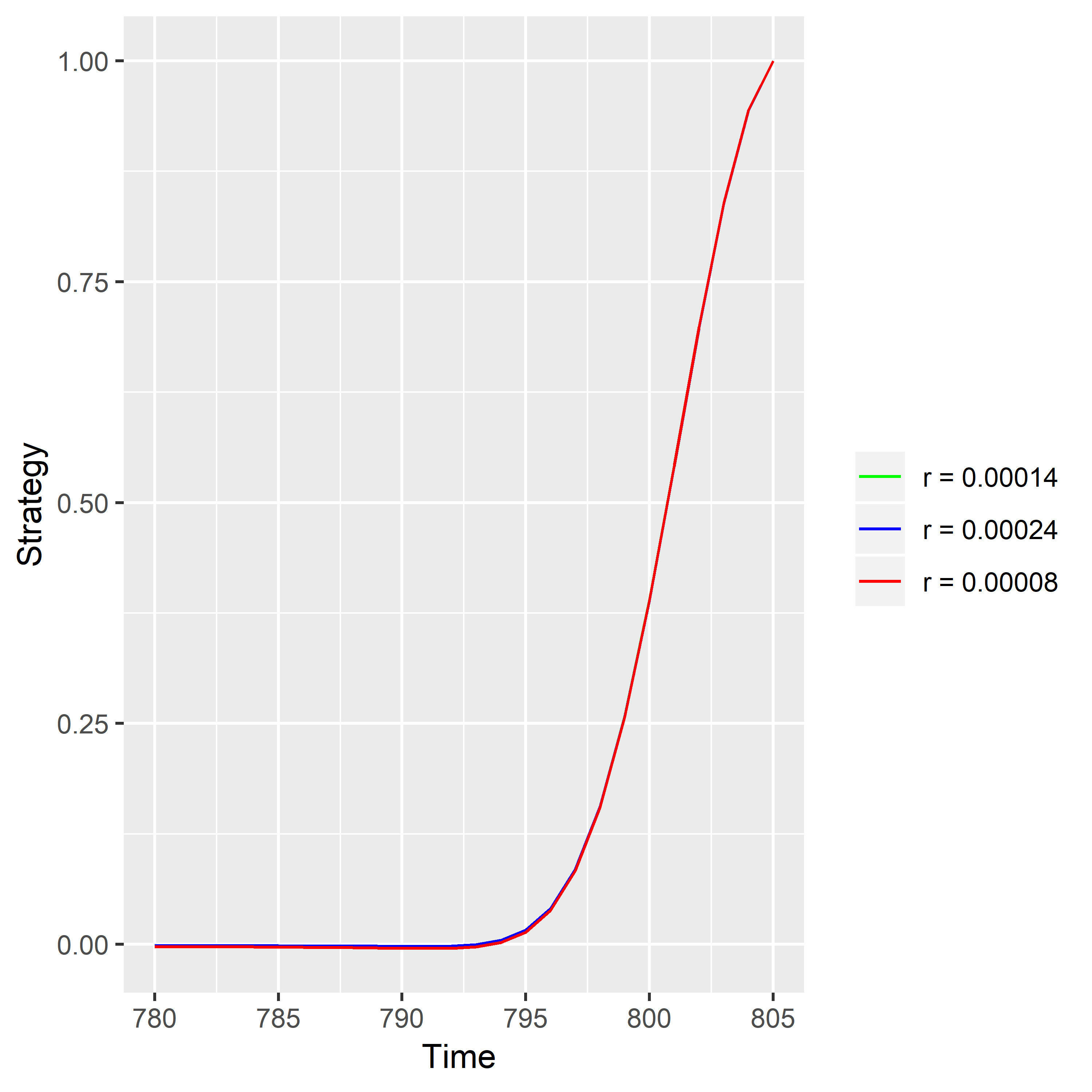}
   \caption{When there is transaction cost }
  \label{fig:4_1}
\end{subfigure}%
\caption{Change in optimal strategy for different values of r when there is transaction cost and no transaction cost considering Pareto distribution}
\label{fig:test}
\end{figure}

\noindent The expression for other values of r are similarly obtained. Plotting \eqref{eq31} for the last 26 days of the historical data we have taken, in R-Studio, we can see how the optimal strategy varies for different values of r. Refer [Figure (\ref{fig:4})] we note that as the time increases from the initial point the fluctuations in $\pi$ are present but as it reaches the terminal point it increases steeply with time. The value it reaches is maximum when the rate of interest is least and is minimum when the rate of interest r is maximum. This is intuitively reasonable as a lower rate of interest would push the investment in the risky asset higher.\newline
\centerline{\textbf{[Figure (\ref{fig:4})] should be placed here}}

\noindent The change in the optimal strategy may or may not  be associated with transaction cost. We have implemented both the scenarios for all the numerical calculations in R-Studio trying to get some interpretation from the graphs.

\subsubsection{Without transaction cost}

\begin{figure}
\centering
\begin{subfigure}{.6\textwidth}
  \centering
  \includegraphics[width=0.95\textwidth]{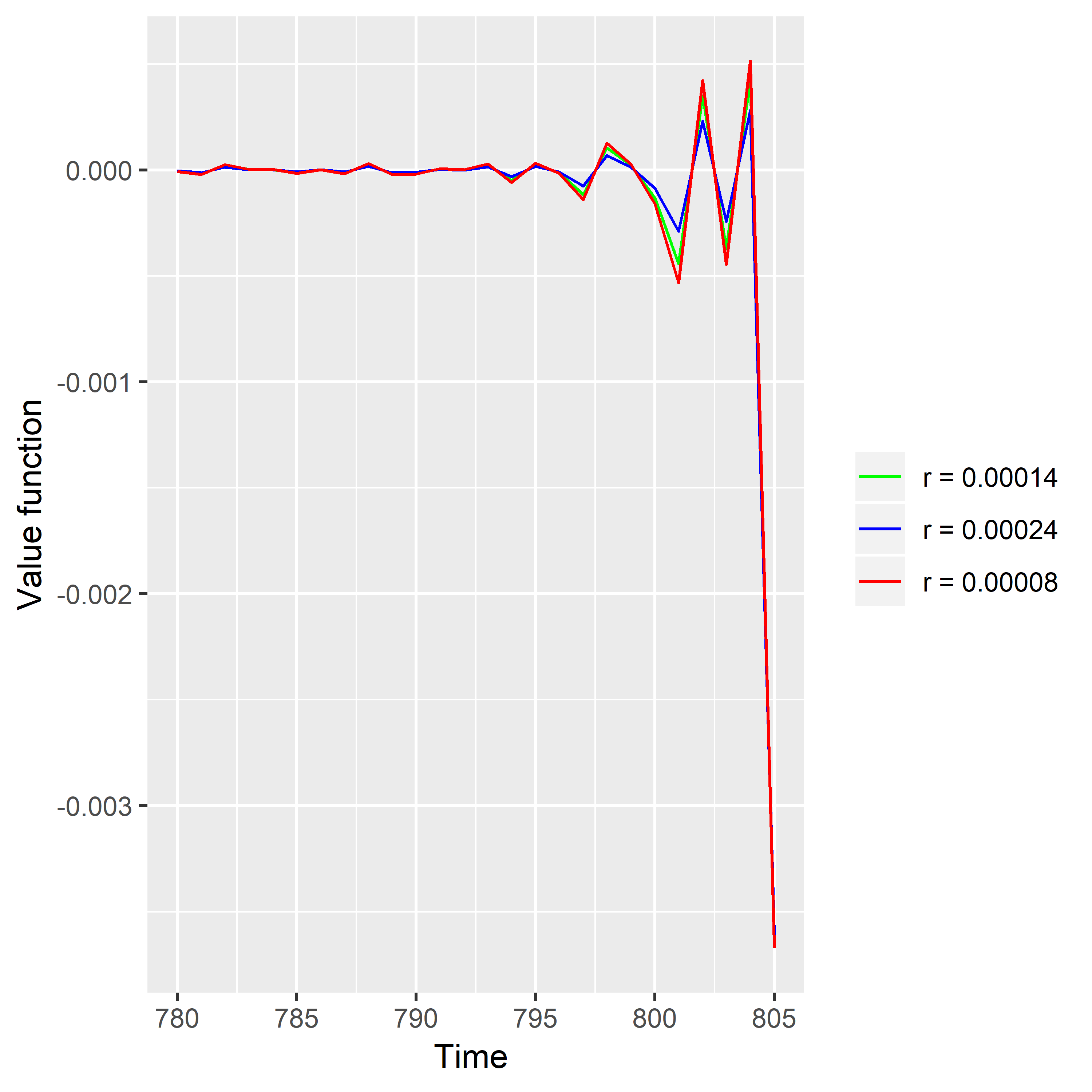}
\caption{When there is no transaction cost }
\label{fig:5}
\end{subfigure}%
\begin{subfigure}{.6\textwidth}
  \centering
  \includegraphics[width=0.95\textwidth]{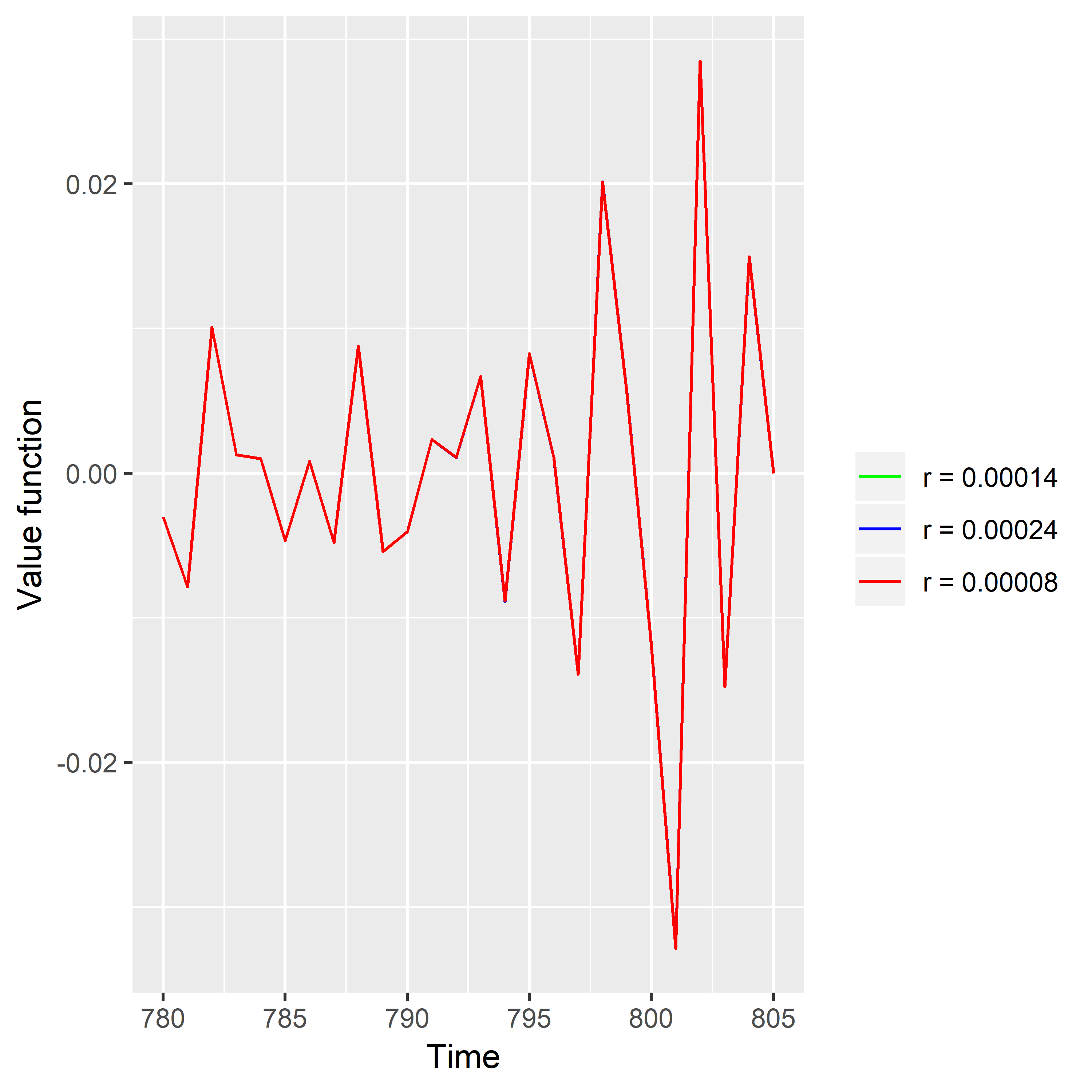}
   \caption{When there is transaction cost }
  \label{fig:5_1}
\end{subfigure}%
\caption{Change in value function for different values of r when there is transaction cost and no transaction cost considering Pareto distribution}
\label{fig:test}
\end{figure}

Putting the values of $\pi$ obtained from \eqref{eq31} into \eqref{eq25} and simultaneously calculating the maximum reward $W_{t-1}$ for the transition from t to t-1 we calculate the value function defined in \eqref{eq28}. If we plot the value function for the last 26 days of the the historical data (Figure (\ref{fig:5})) we can interpret that as it moves towards terminal position the fluctuations in the value functions increases more and for the least rate of interest the fluctuations are maximum and for maximum rate of interest the fluctuations are least.\newline
\centerline{\textbf{[Figure (\ref{fig:5})] should be placed here}}

\begin{figure}
\centering
\begin{subfigure}{.6\textwidth}
  \centering
  \includegraphics[width=0.95\textwidth]{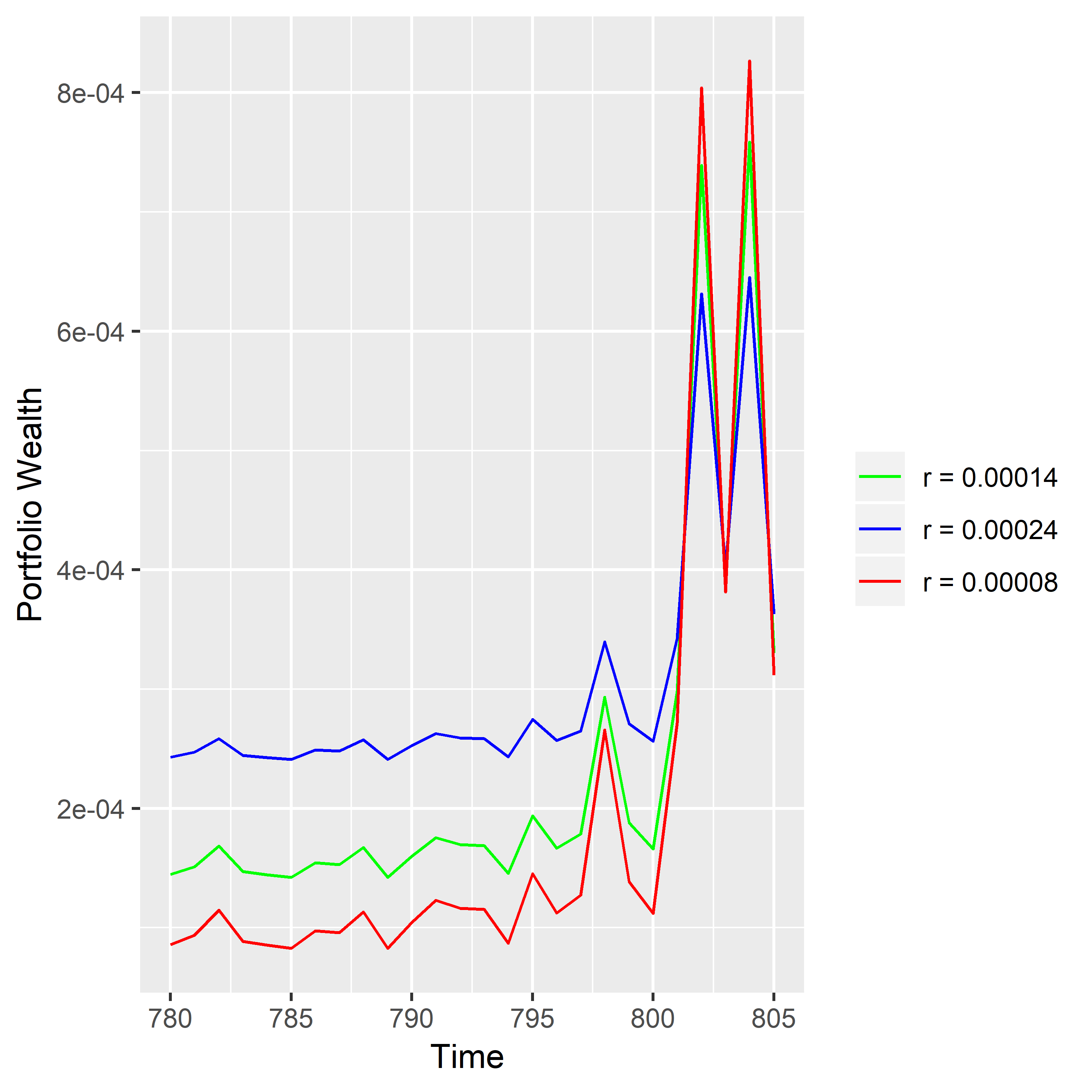}
\caption{When there is no transaction cost }
\label{fig:6}
\end{subfigure}%
\begin{subfigure}{.6\textwidth}
  \centering
  \includegraphics[width=0.95\textwidth]{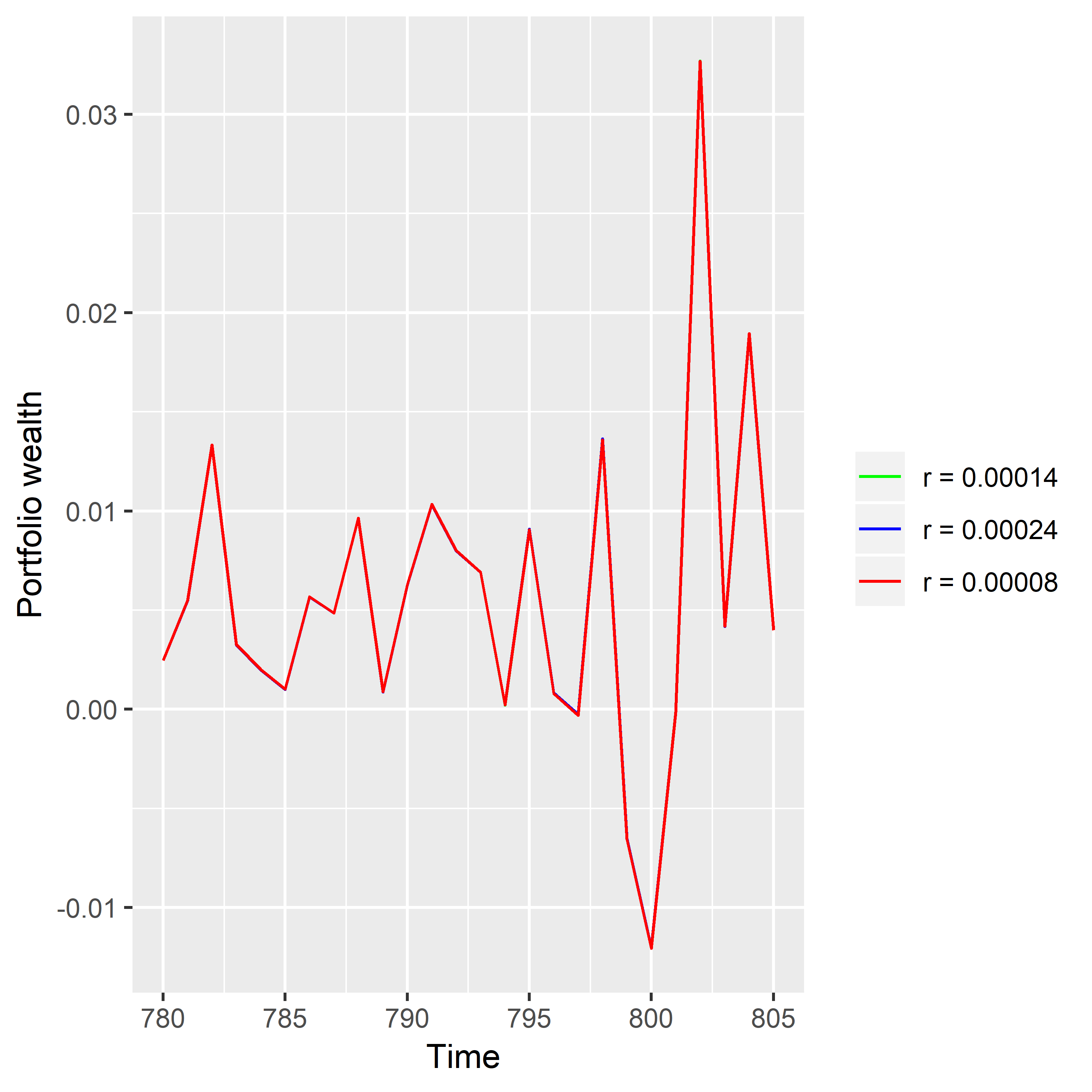}
   \caption{When there is transaction cost }
  \label{fig:6_1}
\end{subfigure}%
\caption{Change in portfolio wealth for different values of r when there is transaction cost and no transaction cost considering Pareto distribution}
\label{fig:test}
\end{figure}

\noindent Finally when we plot the portfolio wealth for the last 26 days of the historical data it can be interpreted that the wealth gradually increases and decreases as it approaches the terminal wealth but before attaining the terminal position the value of the portfolio wealth increases and then decreases. For maximum rate of interest the portfolio wealth is maximum and for least rate on interest the portfolio wealth is minimum (Figure (\ref{fig:6})). This implies that a safer portfolio (when bank rate is higher) is better.\newline 
\centerline{\textbf{[Figure (\ref{fig:6})] should be placed here}}

\subsubsection{With transaction cost}

When transaction cost is taken into account the wealth equation \eqref{eq25} modifies to

\begin{equation}\label{eq32}
L_t = \pi_{t} S_t + (1 - \pi_{t}) r - (\pi_{t} - \pi_{t-1}) r_1 
\end{equation}
and using \eqref{eq32} in the optimisation exercise, the recursive optimal policy modifies to 
\begin{equation}\label{eq33}
 \pi_{t-2} \ge 0.999 \frac{\pi_{t-1}^{2}}{\pi_t} - 0.0547 \pi_{t-1} - 0.002867 + 0.002868 \frac{\pi_{t-1}}{\pi_t}
\end{equation}
we can see that now there is a difference of order 2 when transaction cost is taken into account, where $r_1$ (Transaction cost rate for a change in optimal strategy) = 10\%.\footnote{In this and all the subsequent cases, we have also studied the effect of a change in the transaction cost rate on our variables of interest. As we could detect no discernible difference, we have not reported those results here.} Putting the values of $\pi$ obtained from \eqref{eq33} into \eqref{eq32} and simultaneously calculating the maximum reward $W_{t-1}$ for the transition from t to t-1 we calculate the value function defined in \eqref{eq28}. When the optimal policy is plotted for the last 26 days of the historical data we can see  that for transaction cost scenario the rate of interest doesn't have any significant impact on the optimal strategy (Figure (\ref{fig:4_1})). The optimal policy is almost constant when transaction cost is considered and is at a lower level as compared to no transaction cost scenario but when it reaches the terminal position the optimal strategy value increases  more than the case of no transaction cost.\newline
\centerline{\textbf{[Figure (\ref{fig:4_1})] should be placed here}}
\noindent If we plot the value function for the last 26 days of the historical data denoted by red line (Figure (\ref{fig:5_1})) we see that the value function in initial position does not change as much as compared to the no transaction cost scenario whereas as it closes towards the terminal position the fluctuations are more in the value function and there is no significant effect due to change in the rate of interest.\newline
\centerline{\textbf{[Figure (\ref{fig:5_1})] should be placed here}}
\noindent Finally when we plot the portfolio wealth over the last 26 days of the historical data it can be inferred that the initial wealth follows the same pattern as the no transaction cost scenario but as it approaches the terminal position the portfolio wealth decreases and then increases. There is no significant effect due to change in the rate of interest (Figure (\ref{fig:6_1})). If we compare this with the case of no transaction cost, the fluctuations are more in case of transaction cost scenario.\newline
\centerline{\textbf{[Figure (\ref{fig:6_1})] should be placed here}}

\subsection{Weibull Distribution}

Weibull distribution is used next for modelling the financial return distribution, for the purpose of forecasting tail risk measures. The Weibull performs favourably for VaR forecasting compared to other distributions. The probability density function of a Weibull random variable is,
\begin{equation}
f(x;\lambda ,\alpha)={\begin{cases}{\frac {\alpha}{\lambda }}\left({\frac {x}{\lambda }}\right)^{\alpha-1}e^{-(x/\lambda )^{\alpha}}&x\geq 0,\\0 & x<0,\end{cases}} \nonumber
\end{equation}
where $\alpha$ > 0 is the shape parameter and $\lambda$ > 0 is the scale parameter of the distribution. We fit the Weibull distribution to the data-set and estimate the scale and shape parameters of the distribution using Maximum Likelihood Estimation in the following way:\newline
\noindent The maximum likelihood estimator for the parameter $\lambda $ given $\alpha$ is
\begin{equation}
{\widehat {\lambda }}^{\alpha}={\frac {1}{n}}\sum _{i=1}^{n}x_{i}^{\alpha}\nonumber
\end{equation}
The maximum likelihood estimator for $\alpha$ is the solution for $\alpha$ in the following equation
\begin{equation}
0=\frac {\sum _{i=1}^{n}x_{i}^{\alpha}\ln x_{i}}{\sum _{i=1}^{n}x_{i}^{\alpha}}-{\frac {1}{\alpha}}-{\frac {1}{n}}\sum _{i=1}^{n}\ln x_{i}\nonumber
\end{equation}
This equation defines $\widehat {\alpha}$ only implicitly, one must generally solve for $\alpha$ by numerical means.
When $ x_{1}>x_{2}>\cdots >x_{N}$ are the N largest observed samples from a data-set of more than N samples, then the maximum likelihood estimator for the $\lambda$  parameter given $\alpha$ is
\begin{equation}
{\widehat {\lambda }}^{\alpha}={\frac {1}{N}}\sum _{i=1}^{N}(x_{i}^{\alpha}-x_{N}^{\alpha})\nonumber
\end{equation}
Also given that condition, the maximum likelihood estimator for $\alpha$ is
\begin{equation}
0=\frac {\sum _{i=1}^{N}(x_{i}^{\alpha}\ln x_{i}-x_{N}^{\alpha}\ln x_{N})}{\sum _{i=1}^{N}(x_{i}^{\alpha}-x_{N}^{\alpha})}-{\frac {1}{N}}\sum _{i=1}^{N}\ln x_{i}\nonumber
\end{equation}
Again, this being an implicit function, one must generally solve for $\alpha$ by numerical means.
Once we get the parameters estimated, we calculate the quantile $Q_{0.05}$ for the distribution. The constraint equation is again similar to \eqref{eq29}.
The scale and shape parameters are as follows (refer Table \ref{weibull}).\newline
\centerline{\textbf{Table \ref{weibull} should be here}}

\noindent Again, once the parameters are estimated using 700 return data points, we check whether the next 50 data points are coming from the same distribution using the one-sample Kolmogorov-Smirnov test.
We obtain the $D_{50} = 0.152271$ which is less than the $D_{50,0.05} = 0.19206$. Since $D_{50} < D_{50,0.05}$, we cannot reject the null hypothesis and conclude that there is no significant difference between the data and data coming from an Weibull distribution (with $\lambda$ = 1.2630 and $\alpha$ = 0.0104).\newline

\begin{figure}
\centering
\begin{subfigure}{.6\textwidth}
  \centering
  \includegraphics[width=0.95\textwidth]{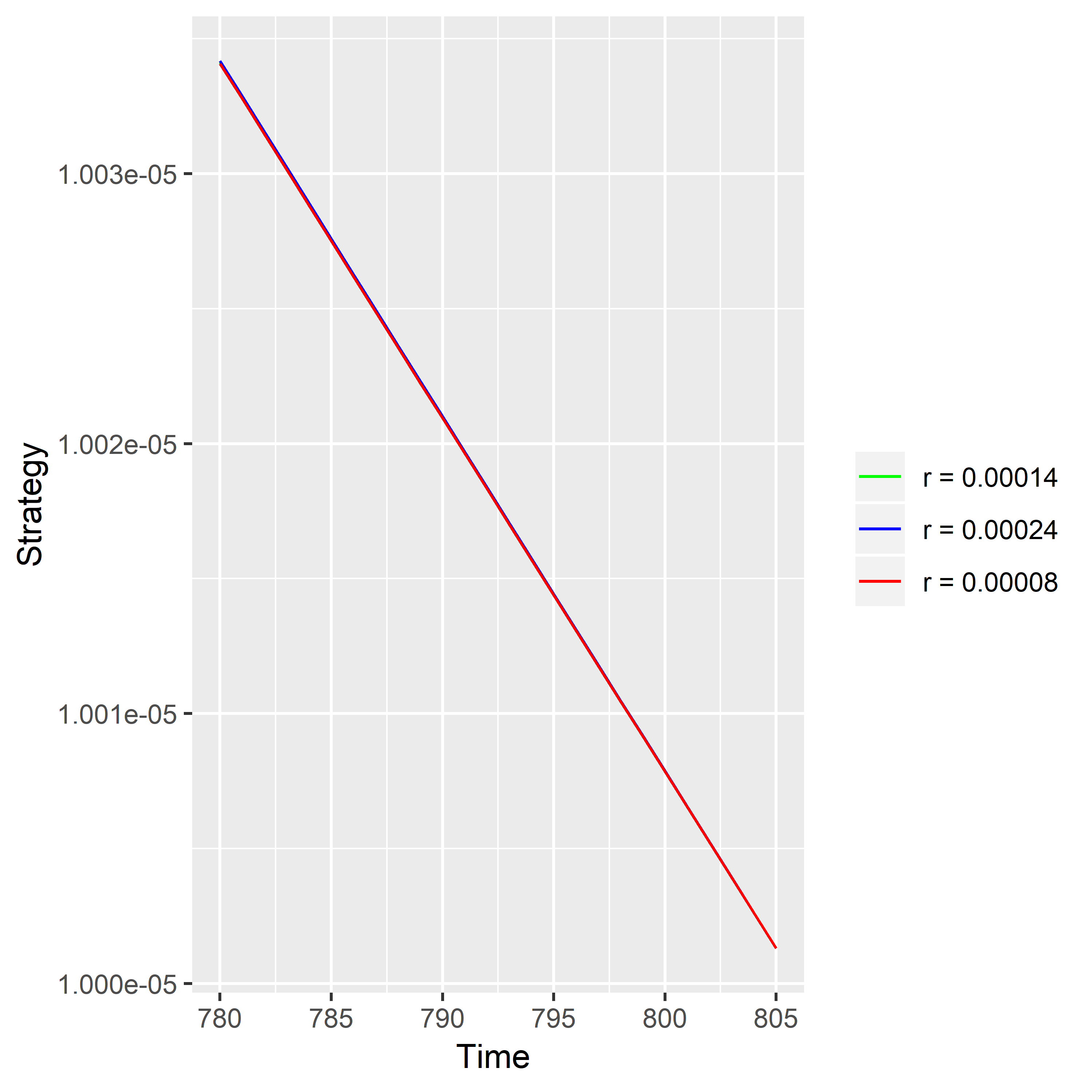}
\caption{When there is no transaction cost }
\label{fig:7}
\end{subfigure}%
\begin{subfigure}{.6\textwidth}
  \centering
  \includegraphics[width=0.95\textwidth]{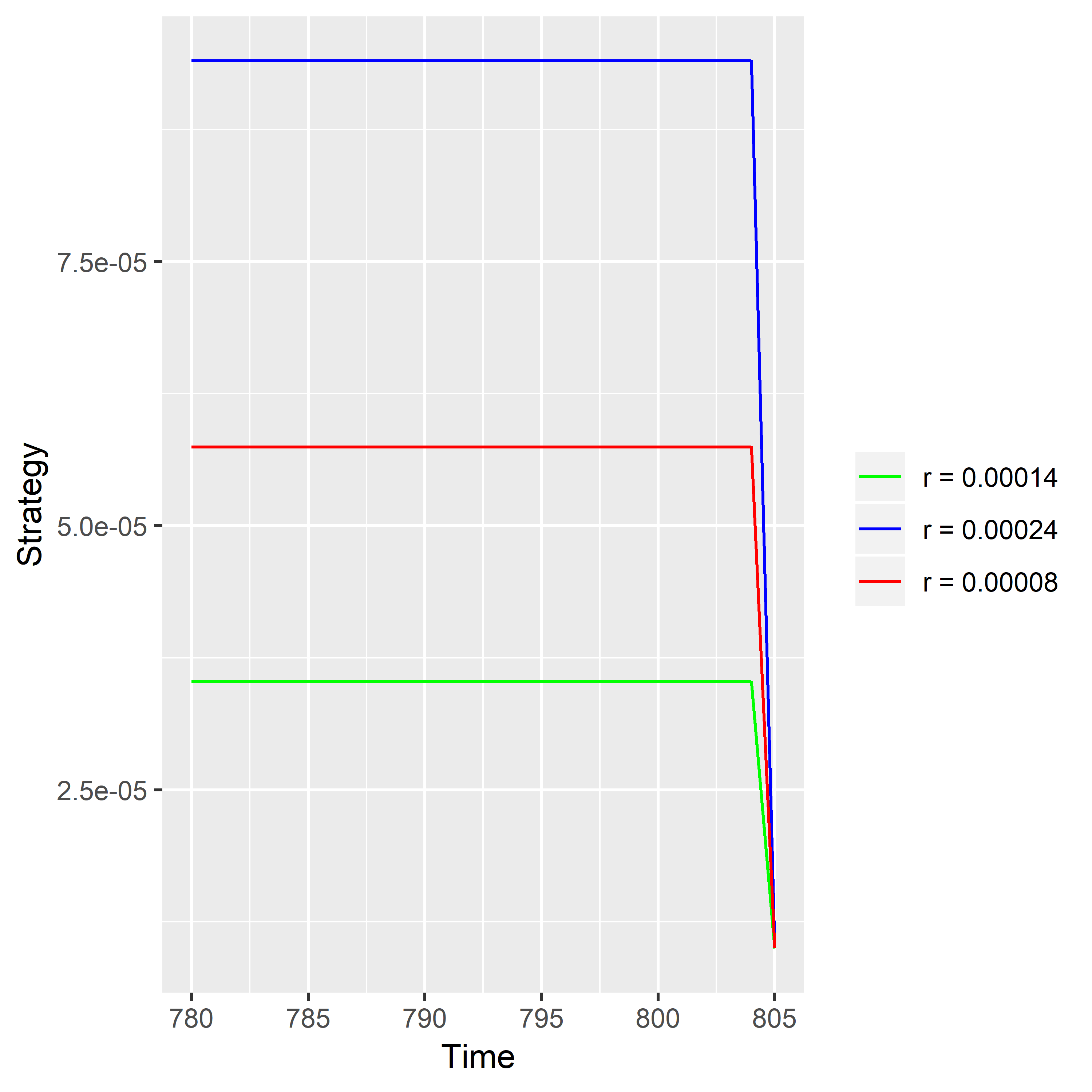}
   \caption{When there is transaction cost }
  \label{fig:7_1}
\end{subfigure}%
\caption{Change in optimal strategy for different values of r when there is transaction cost and no transaction cost considering Weibull distribution}
\label{fig:test}
\end{figure}

\noindent The rate of interest (daily) is taken to be one of three alternative values as for Pareto. Putting the value of the quantile in equation \eqref{eq29} and putting the value of wealth from \eqref{eq25} into the same equation we get,

\begin{equation}
  \frac{ P (\pi_{t-1}S_{t-1} + (M-\pi_{t-1})r > 0.00078\lambda, \pi_{t}S_{t} + (M-\pi_{t})r > 0.00078\lambda)}{P (\pi_{t}S_{t} + (M-\pi_{t})r > 0.00078\lambda)} \ge 0.95 \nonumber 
\end{equation}
Taking M = 1,
\begin{equation}\
  \frac{ P \big(S_{t-1} > \frac{0.00078\lambda + r(\pi_{t-1}-1)}{\pi_{t-1}}, S_{t} > \frac{0.00078\lambda + r(\pi_{t}-1)}{\pi_{t}}\big)}{P \big(S_{t} > \frac{0.00078\lambda + r(\pi_{t}-1)}{\pi_{t}}\big)} \ge 0.95 \nonumber
\end{equation}
For simplicity let us denote the upper and lower limits by the following constants
\begin{center}
\begin{tabular}{c c l l}
     $w_t$ &=& $\frac{0.00078\lambda + r(\pi_{t}-1)}{\pi_{t}}$ 
\end{tabular}
\end{center}
Then the constraint becomes
\begin{equation} \label{eq34}
    \frac{\int\limits_{w_{t-1}}^{w_t} \big(\frac{0.0104}{1.2630}\big)\big(\frac{x}{1.2630}\big)^{0.0104 - 1}\big(e^{\frac{-x}{1.2630}}\big)^{0.0104} dx}{1-\int\limits_{0}^{w_t} \big(\frac{0.0104}{1.2630}\big)\big(\frac{x}{1.2630}\big)^{0.0104 - 1}\big(e^{\frac{-x}{1.2630}}\big)^{0.0104} dx} \ge 0.95
\end{equation}
Solving \eqref{eq33} and putting the values of $\lambda$ and $r = 0.00014$ we get the recursive relation between the optimal strategies for t to t-1,
\begin{equation}\label{eq35}
  \pi_{t-1} \ge \frac{\pi_t}{(1 - 13.08\pi_t)}  
\end{equation}
We have assumed the terminal optimal point be 0.00001. Plotting \eqref{eq35} for the last 26 days of the historical data in R-Studio we can see how the optimal policy varies (Figure (\ref{fig:7})) and we notice that as the time increases from the initial point the change in $\pi$ is also linear with time and is same for all the rates of interest r.\newline
\centerline{\textbf{[Figure (\ref{fig:7})] should be placed here}}
\noindent Again, to check the association between the optimal strategy and transaction cost, we have implemented both the scenarios. 

\subsubsection{Without transaction cost}

\begin{figure}
\centering
\begin{subfigure}{.6\textwidth}
  \centering
  \includegraphics[width=0.95\textwidth]{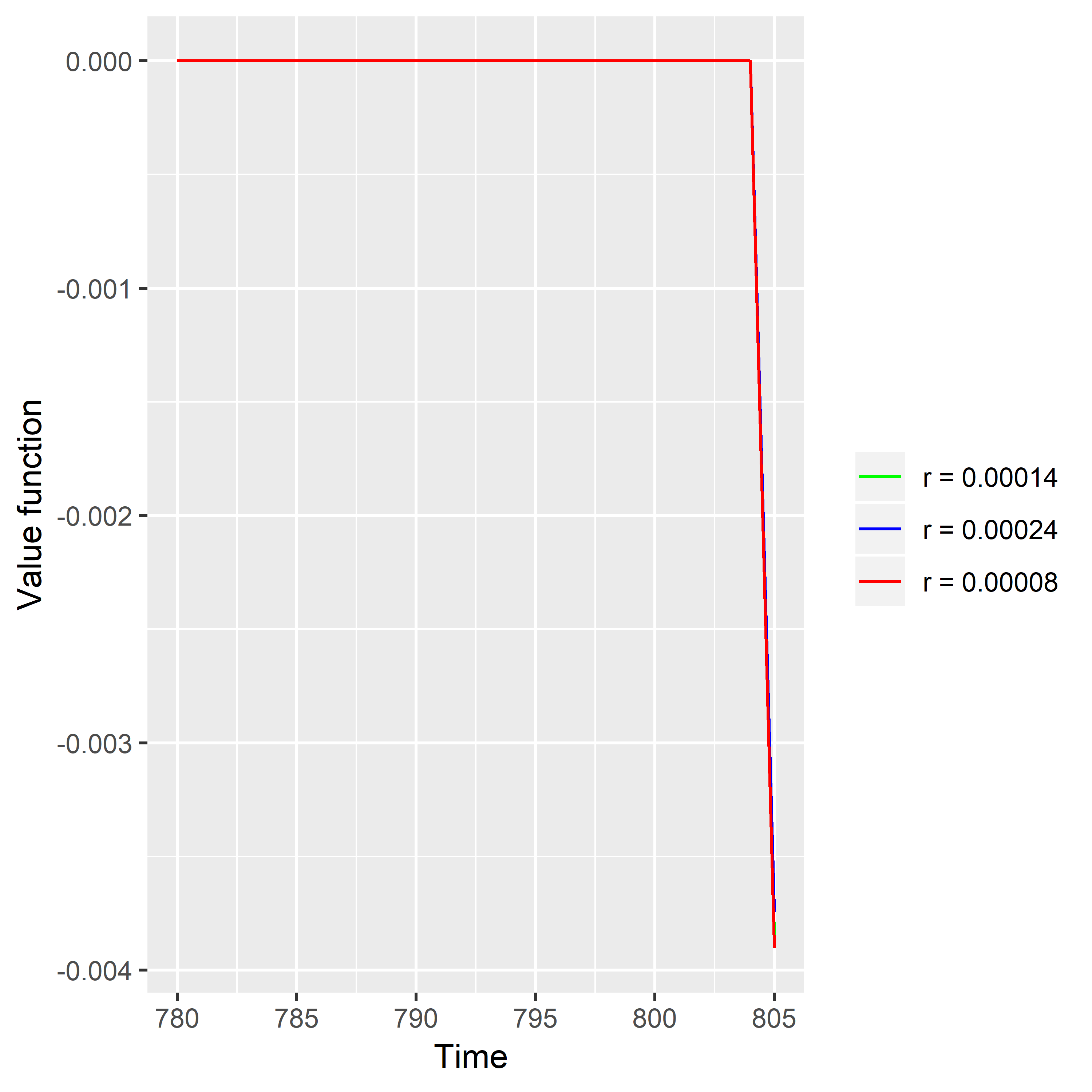}
\caption{When there is no transaction cost }
\label{fig:8}
\end{subfigure}%
\begin{subfigure}{.6\textwidth}
  \centering
  \includegraphics[width=0.95\textwidth]{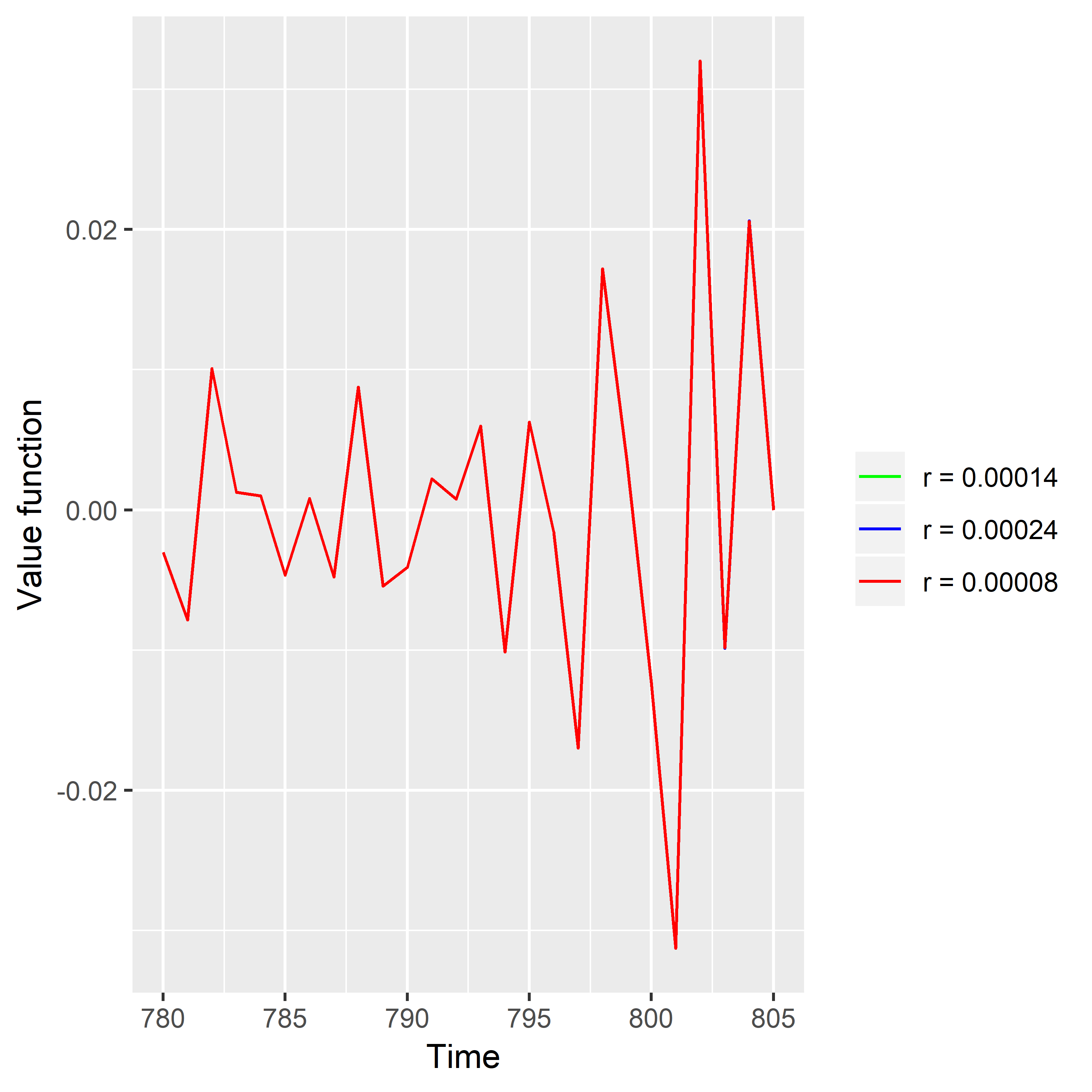}
   \caption{When there is transaction cost }
  \label{fig:8_1}
\end{subfigure}%
\caption{Change in value function for different values of r when there is transaction cost and no transaction cost considering Weibull distribution}
\label{fig:test}
\end{figure}

Putting the values of $\pi$ obtained from \eqref{eq35} into \eqref{eq25} and simultaneously calculating the maximum reward $W_{t-1}$ for the transition from t to t-1 we calculate the value function defined in \eqref{eq28}. If we plot the value function for the last 26 days of the historical data (Figure (\ref{fig:8})) we can interpret that initially as the time period moves the value function is constant but once it is close to terminal position the fluctuation decreases. Overall the value function doesn't depend on the rate of interest r.\newline
\centerline{\textbf{[Figure (\ref{fig:8})] should be placed here}}

\begin{figure}
\centering
\begin{subfigure}{.6\textwidth}
  \centering
  \includegraphics[width=0.95\textwidth]{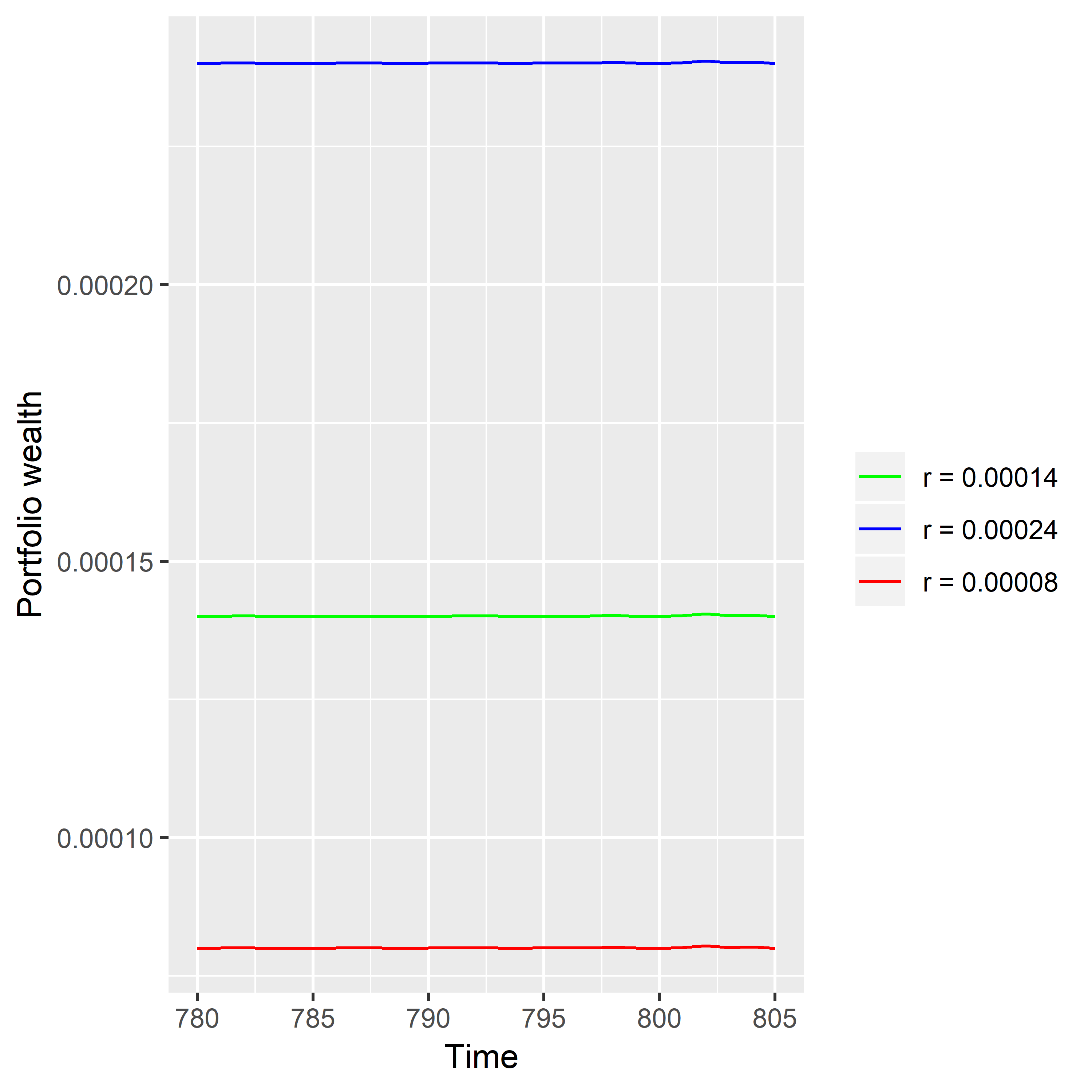}
\caption{When there is no transaction cost }
\label{fig:9}
\end{subfigure}%
\begin{subfigure}{.6\textwidth}
  \centering
  \includegraphics[width=0.95\textwidth]{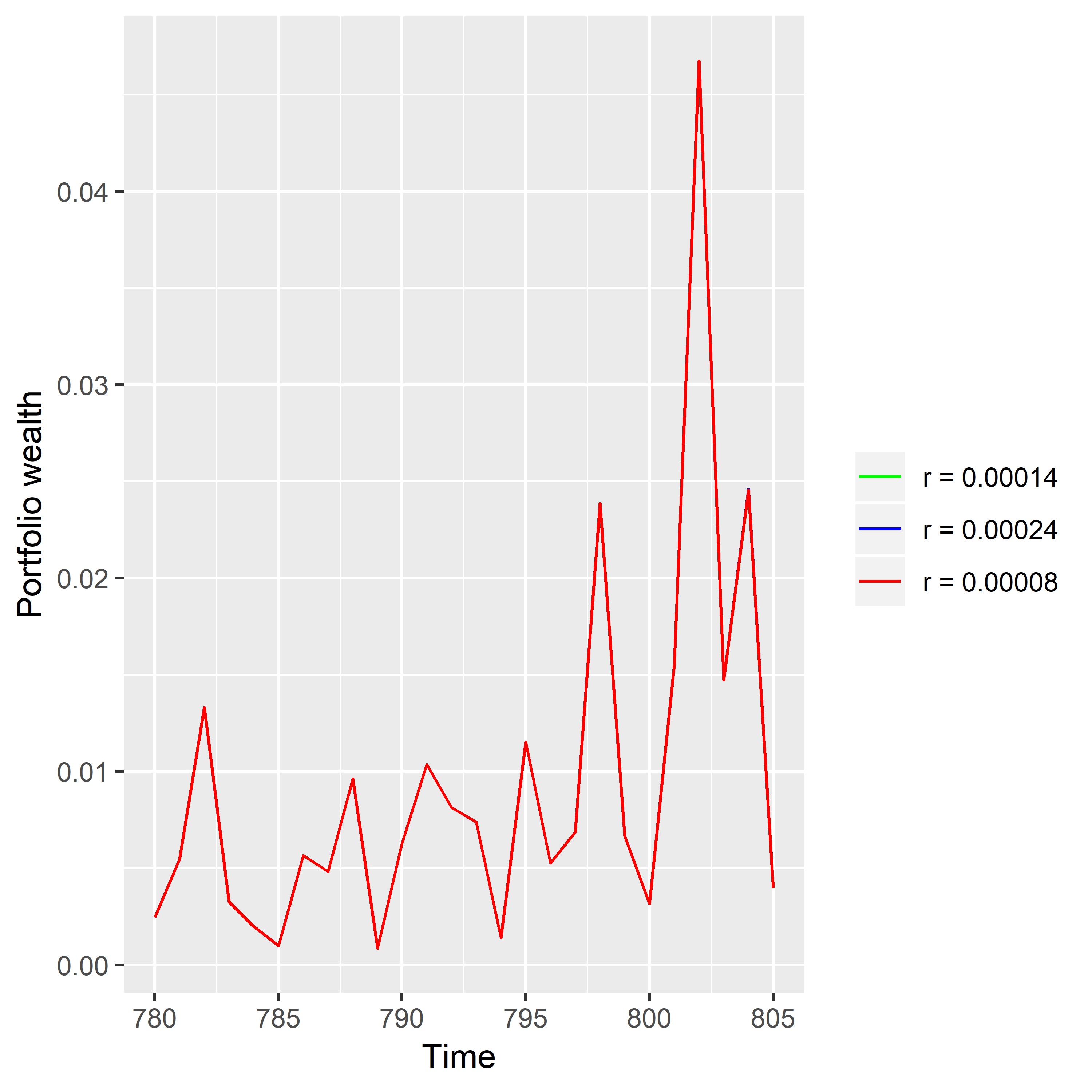}
   \caption{When there is transaction cost }
  \label{fig:9_1}
\end{subfigure}%
\caption{Change in portfolio wealth for different values of r when there is transaction cost and no transaction cost considering Weibull distribution}
\label{fig:test}
\end{figure}

\noindent Finally when we plot the portfolio wealth over the last 26 days of the historical data it is observed that throughout the time period, portfolio wealth fluctuates very less but when the rate of interest r is maximum then the portfolio wealth is also maximum and when the rate of interest is least then the portfolio wealth is minimum (Figure (\ref{fig:9})).\newline
\centerline{\textbf{[Figure (\ref{fig:9})] should be placed here}}

\subsubsection{With transaction cost}

When transaction cost is taken into account the wealth equation \eqref{eq25} modifies to
\begin{equation}\label{eq36}
L_t = \pi_{t} S_t + (1 - \pi_{t}) r - (\pi_{t} - \pi_{t-1}) r_1 
\end{equation}
and using \eqref{eq36} the recursive optimal policy modifies to
\begin{equation}\label{eq37}
    \pi_{t-2} = \frac{\pi_{t-1}^{2}}{\pi_{t}} + \frac{\pi_{t-1}}{1.958 \pi_{t}} - 0.5107 - 6.680 \pi_{t-1}
\end{equation}
The recursive equation is of order 2 and $r_1$ (Transaction cost rate for a change in optimal strategy) = 10\%. Putting the values of $\pi$ obtained from \eqref{eq37} into \eqref{eq36} and simultaneously calculating the maximum reward $W_{t-1}$ for the transition from t to t-1 we calculate the value function defined in \eqref{eq28}. When the optimal policy is plotted for the last 26 days of the historical data it is observed that the optimal policy remains constant at a particular value for a period of time but as it approaches the terminal position the asset needs to be liquidated. Here the rate of interest r has a clear effect on the strategy (Figure (\ref{fig:7_1})). For different rate of interest the liquidation of the asset is different. In case of no transaction cost the rate of interest had no effect on the optimal strategy.\newline
\centerline{\textbf{[Figure (\ref{fig:7_1})] should be placed here}}
 
\noindent If we plot the value function over the last 26 days of the historical data (Figure (\ref{fig:8_1})) we can perceive that the value function keeps fluctuating and there is no effect of the rate of interest.\newline
\centerline{\textbf{[Figure (\ref{fig:8_1})] should be placed here}}

\noindent Finally  when  we  plot  the  portfolio  wealth  over the last 26 days of the historical data it can be interpreted that the wealth fluctuates throughout the time period and there is no significant difference while changing the rate of interest (Figure (\ref{fig:9_1})). We obtain a different result during the no transaction cost scenario.\newline
\centerline{\textbf{[Figure (\ref{fig:9_1})] should be placed here}}

\subsection{Inverse Gaussian Distribution}

In probability theory, the inverse Gaussian (IG) distribution (also known as the Wald distribution) is a two-parameter family of continuous probability distributions with support on $(0,\infty)$. Its probability density function is given by
\begin{equation}
f(x;\mu ,\lambda )={\sqrt {\frac {\lambda }{2\pi x^{3}}}}\exp \left[-{\frac {\lambda (x-\mu )^{2}}{2\mu ^{2}x}}\right]\nonumber
\end{equation}
for x > 0, where $ \mu >0$ is the mean and $ \lambda >0$ is the shape parameter. The parameters are calculated using Maximum Likelihood Estimator as follows:
\begin{equation}
X_{i}\sim \operatorname {IG} (\mu ,\lambda w_{i}),\,\,\,\,\,\,i=1,2,\ldots ,n \nonumber
\end{equation}
with all $w_i$ known, $(\mu, \lambda)$ unknown and all $X_i$ independent has the following likelihood function
\begin{equation}
 L(\mu ,\lambda )=\left({\frac {\lambda }{2\pi }}\right)^{\frac {n}{2}}\left(\prod _{i=1}^{n}{\frac {w_{i}}{X_{i}^{3}}}\right)^{\frac {1}{2}}\exp \left({\frac {\lambda }{\mu }}\sum _{i=1}^{n}w_{i}-{\frac {\lambda }{2\mu ^{2}}}\sum _{i=1}^{n}w_{i}X_{i}-{\frac {\lambda }{2}}\sum _{i=1}^{n}w_{i}{\frac {1}{X_{i}}}\right) \nonumber
\end{equation}
Solving the maximum likelihood equation yields the following maximum likelihood estimates
\begin{equation}
{\widehat {\mu }}={\frac {\sum _{i=1}^{n}w_{i}X_{i}}{\sum _{i=1}^{n}w_{i}}},\,\,\,\,\,\,\,\,{\frac {1}{\widehat {\lambda }}}={\frac {1}{n}}\sum _{i=1}^{n}w_{i}\left({\frac {1}{X_{i}}}-{\frac {1}{\widehat {\mu }}}\right)\nonumber
\end{equation}
${\widehat {\mu }}$ and ${\widehat {\lambda }}$ are independent and
\begin{equation}
{\widehat {\mu }}\sim \operatorname {IG} \left(\mu ,\lambda \sum _{i=1}^{n}w_{i}\right),\qquad {\frac {n}{\widehat {\lambda }}}\sim {\frac {1}{\lambda }}\chi _{n-1}^{2}\nonumber
\end{equation}
Fitting an Inverse Gaussian distribution to the data-set and estimating the mean and shape parameters of the distribution, we calculate the quantile $Q_{0.05}$ for the distribution. The constraint equation is again similar to \eqref{eq29} and comparison can be made of the optimal strategy that we have obtained for Pareto and Weibull distribution in the previous sections.\newline
The mean and shape parameters obtained are as follows (refer Table \ref{invgauss}).\newline
\centerline{\textbf{Table \ref{invgauss} should be here}} 

\noindent Once the parameters are estimated using 700 return data points, we check whether the next 50 data points are coming from the same distribution. We obtain the value for the Kolmogorov-Smirnov statistic $D_{50} = 0.14623$ which is less than the $D_{50,0.05} = 0.19206$. Since $D_{50} < D_{50,0.05}$, we cannot reject the null hypothesis and conclude that there is no significant difference between the data and the Inverse Gaussian distribution (with $\mu$ = 0.0097 and $\lambda$ = 0.0044).\newline

\begin{figure}
\centering
\begin{subfigure}{.6\textwidth}
  \centering
  \includegraphics[width=0.95\textwidth]{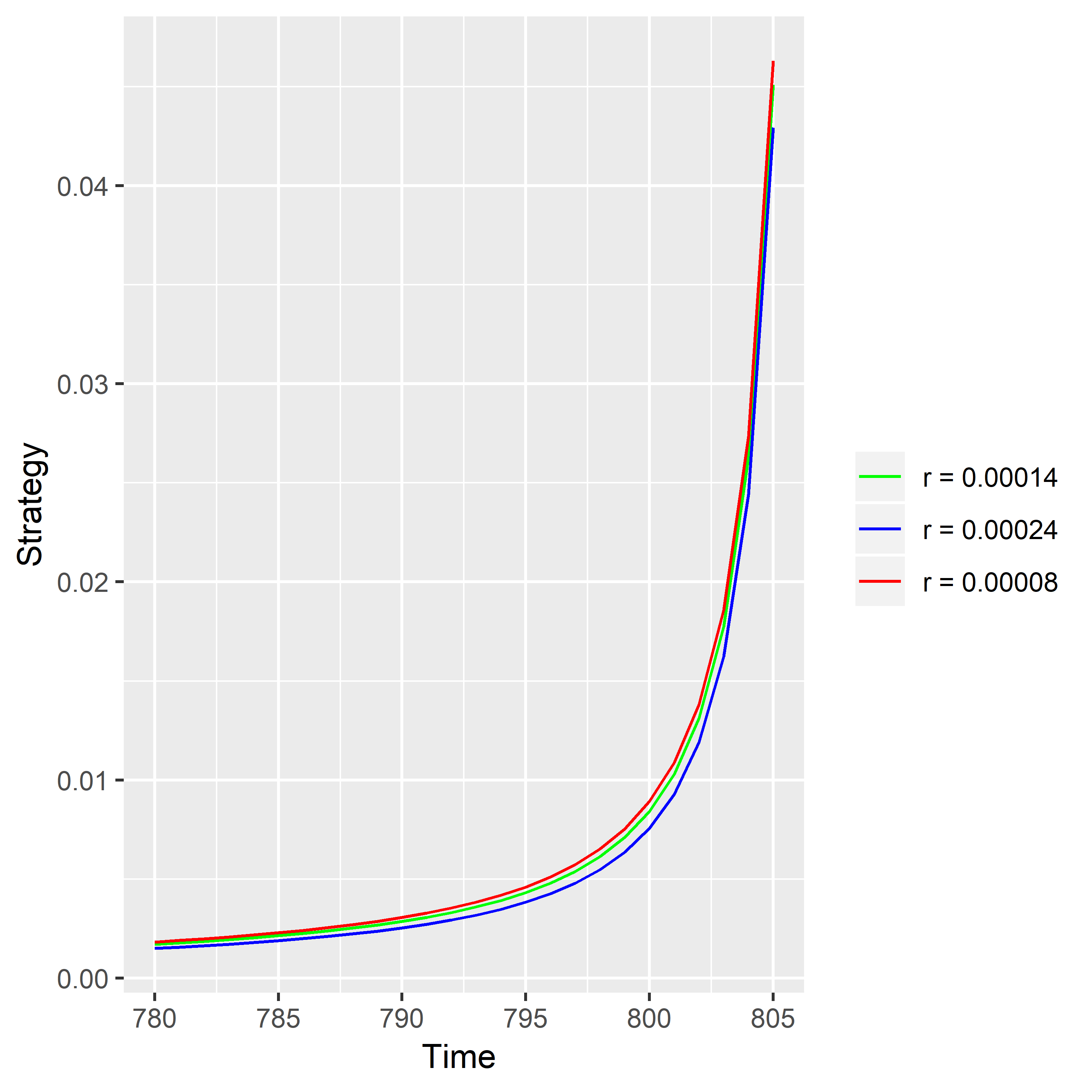}
\caption{When there is no transaction cost }
\label{fig:34}
\end{subfigure}%
\begin{subfigure}{.6\textwidth}
  \centering
  \includegraphics[width=0.95\textwidth]{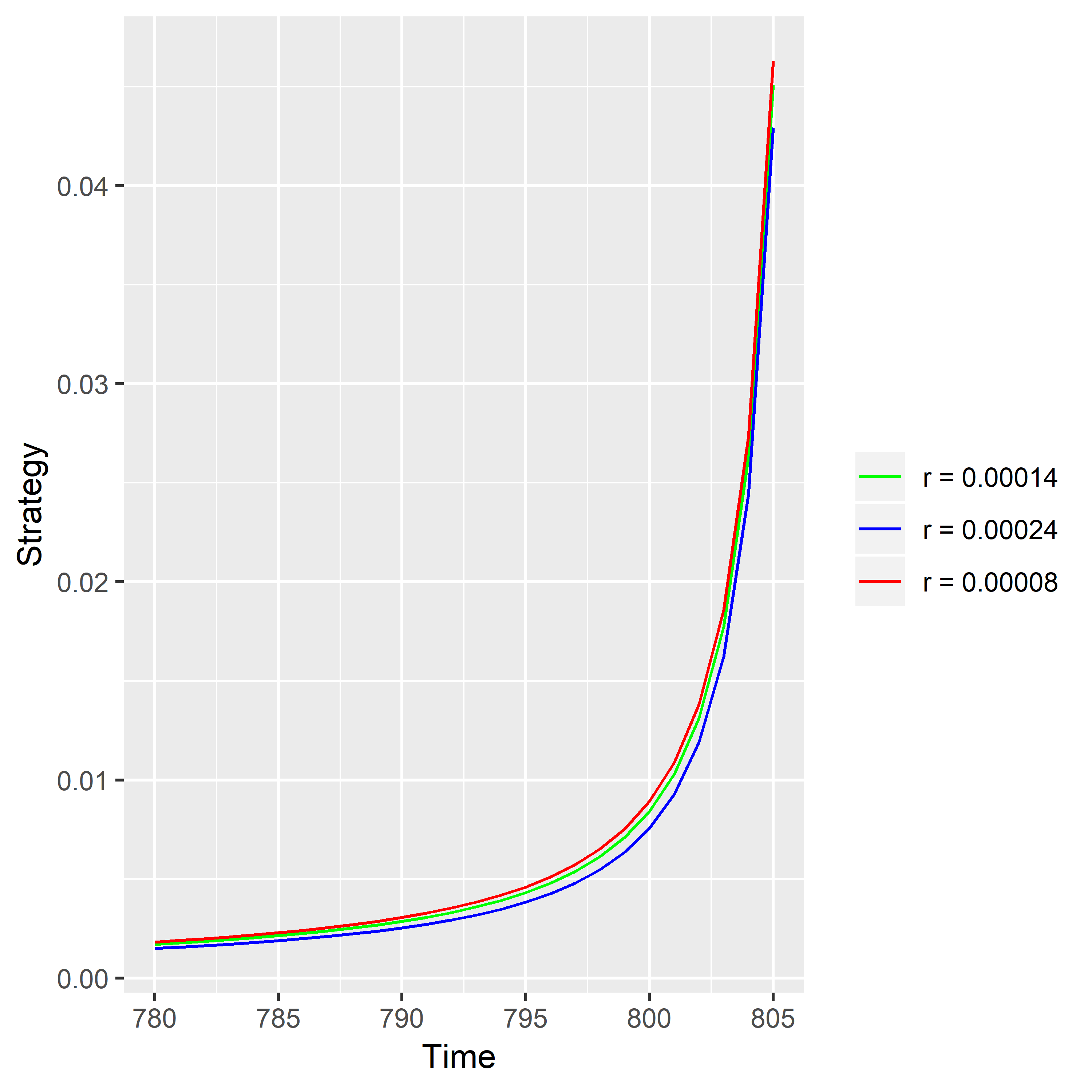}
   \caption{When there is transaction cost }
  \label{fig:34_1}
\end{subfigure}%
\caption{Change in optimal strategy for different values of r when there is transaction cost and no transaction cost considering Weibull distribution}
\label{fig:test}
\end{figure}

\noindent We again consider three alternative rates of interest. Putting the value of the quantile in equation \eqref{eq29} and putting the value of wealth from \eqref{eq25} in to the same equation we get,

\begin{equation}
  \frac{ P (\pi_{t-1}S_{t-1} + (M-\pi_{t-1})r > 0.00096, \pi_{t}S_{t} + (M-\pi_{t})r > 0.00096)}{P (\pi_{t}S_{t} + (M-\pi_{t})r > 0.00096)} \ge 0.95 \nonumber 
\end{equation}
Taking M = 1,
\begin{equation}
  \frac{ P \big(S_{t-1} > \frac{0.00096 + r(\pi_{t-1}-1)}{\pi_{t-1}}, S_{t} > \frac{0.00096 + r(\pi_{t}-1)}{\pi_{t}}\big)}{P \big(S_{t} > \frac{0.00096 + r(\pi_{t}-1)}{\pi_{t}}\big)} \ge 0.95 \nonumber
\end{equation}

\begin{eqnarray}
  1.95\Bigg(\Phi\Bigg(\sqrt{\frac{\lambda}{\frac{0.00096 + r(\pi_{t}-1)}{\pi_{t}}}}\Bigg(\frac{\frac{0.00096 + r(\pi_{t}-1)}{\pi_{t}}}{\mu} - 1\Bigg)\Bigg) + \exp\Bigg({\frac{2\lambda}{\mu}}\Bigg)\Phi\Bigg(-\sqrt{\frac{\lambda}{\frac{0.00096 + r(\pi_{t}-1)}{\pi_{t}}}}\Bigg(\frac{\frac{0.00096 + r(\pi_{t}-1)}{\pi_{t}}}{\mu} + 1\Bigg)\Bigg)\Bigg) - \nonumber \\ \Bigg( \Phi\Bigg(\sqrt{\frac{\lambda}{\frac{0.00096 + r(\pi_{t-1}-1)}{\pi_{t-1}}}}\Bigg(\frac{\frac{0.00096 + r(\pi_{t-1}-1)}{\pi_{t-1}}}{\mu} - 1\Bigg)\Bigg) + \exp\Bigg({\frac{2\lambda}{\mu}}\Bigg)\Phi\Bigg(-\sqrt{\frac{\lambda}{\frac{0.00096 + r(\pi_{t-1}-1)}{\pi_{t-1}}}}\Bigg(\frac{\frac{0.00096 + r(\pi_{t-1}-1)}{\pi_{t-1}}}{\mu} + 1\Bigg)\Bigg) \Bigg) \ge 0.95 \nonumber
\end{eqnarray}
For simplicity let us denote the following terms with these symbols
\begin{center}
\begin{tabular}{c c l l}
     $y_{t}$ &=& $\sqrt{\lambda / \Big( \frac{0.00096 + r(\pi_{t}-1)}{\pi_{t}}\Big)},$ \\ $g_t$ &=& $\Big( \frac{0.00096 + r(\pi_{t}-1)}{\pi_{t}} \Big) / \mu.$ 
\end{tabular}
\end{center}
Then, the desired condition can be rewritten as
\begin{eqnarray}
  1.95\Big(\Phi\big(y_t\big(g_t - 1\big)\big) + 2.48\times\Phi\big(-y_t\big(g_t + 1\big)\big)\Big) -  \Big( \Phi\big(y_{t-1}\big(g_{t-1} - 1\big)\big) + 2.48\times\Phi\big(-y_{t-1}\big(g_{t-1} + 1\big)\big) \Big) \ge 0.95. \nonumber
\end{eqnarray}
\begin{eqnarray}\label{ing}
\text{Or, }  1.95\Phi\big(y_t g_t - y_t\big) + 4.84\Phi\big(-y_t g_t - y_t\big) -  \Phi\big(y_{t-1} g_{t-1} - y_{t-1}\big) - 2.48\Phi\big(-y_{t-1} g_{t-1} - y_{t-1}\big) \ge 0.95.
\end{eqnarray}
We have assumed the terminal optimal strategy to be 0.1. Plotting \eqref{ing} for the last 26 days of the historical data in R-Studio with the help of non linear equation solver we can note the variation of the optimal policy (Figure (\ref{fig:34})) and observe that the change in $\pi$ is non-linear with time and is same for all the rates of interest r.\newline
\centerline{\textbf{[Figure (\ref{fig:34})] should be placed here}}

\noindent Again, to check the association between the optimal strategy and transaction cost, we have implemented both the scenarios. 

\subsubsection{Without transaction cost}

\begin{figure}
\centering
\begin{subfigure}{.6\textwidth}
  \centering
  \includegraphics[width=0.95\textwidth]{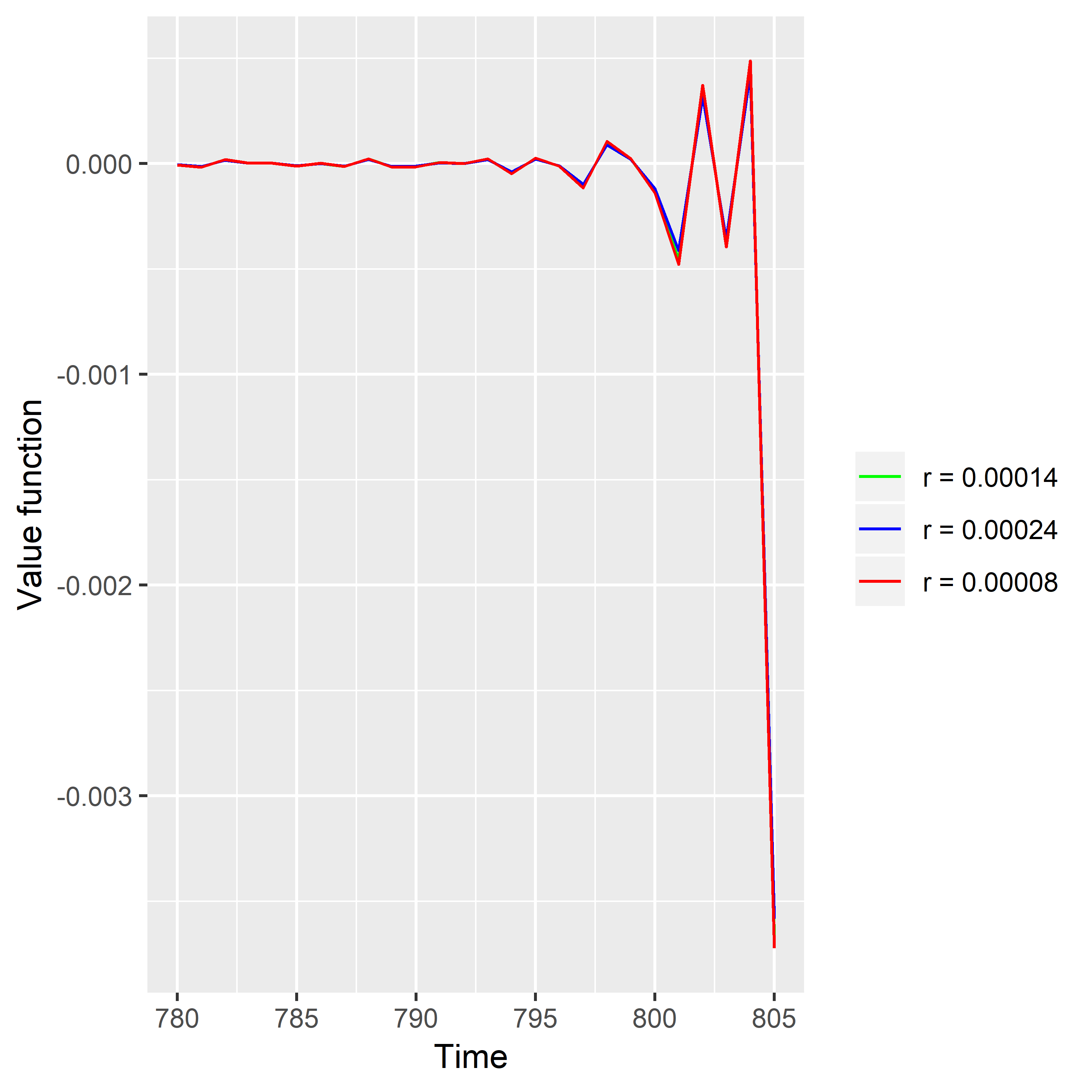}
\caption{When there is no transaction cost }
\label{fig:35}
\end{subfigure}%
\begin{subfigure}{.6\textwidth}
  \centering
  \includegraphics[width=0.95\textwidth]{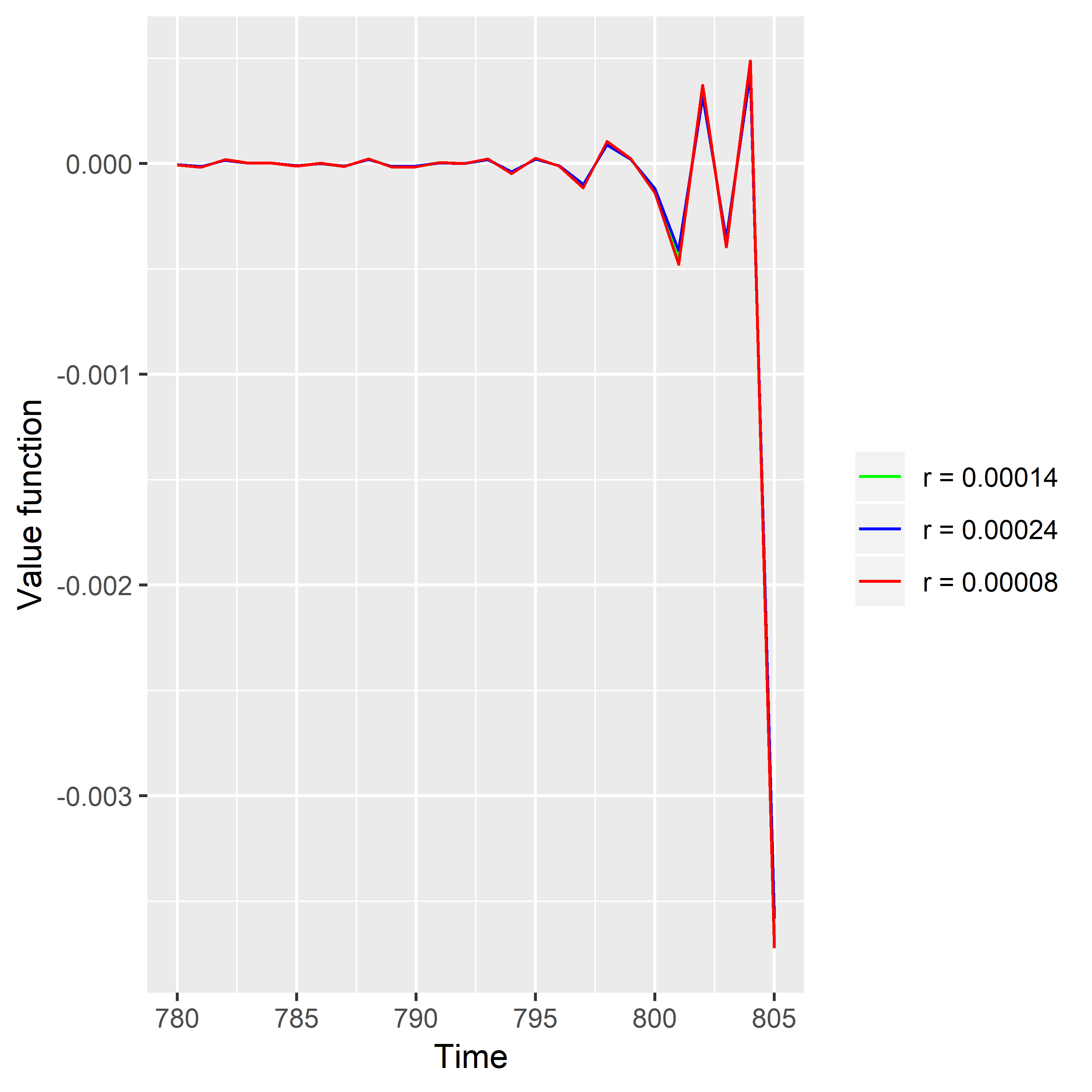}
   \caption{When there is transaction cost }
  \label{fig:35_1}
\end{subfigure}%
\caption{Change in value function for different values of r when there is transaction cost and no transaction cost considering Weibull distribution}
\label{fig:test}
\end{figure}

Putting the values of $\pi$ obtained from \eqref{ing} into \eqref{eq25} and simultaneously calculating the maximum reward $W_{t-1}$ for the transition from t to t-1, we calculate the value function defined in \eqref{eq28}. If we plot the value function for the last 26 days of the historical data (Figure (\ref{fig:35})) we can interpret that with gradual increase in time the value function fluctuates slowly but when it approaches the terminal position the fluctuation increases. Overall the value function depend slightly on the rate of interest r. When the rate of interest is least then the values obtained by the value function is minimum and for maximum rate of interest the value is maximum.\newline
\centerline{\textbf{[Figure (\ref{fig:35})] should be placed here}}

\begin{figure}
\centering
\begin{subfigure}{.6\textwidth}
  \centering
  \includegraphics[width=0.95\textwidth]{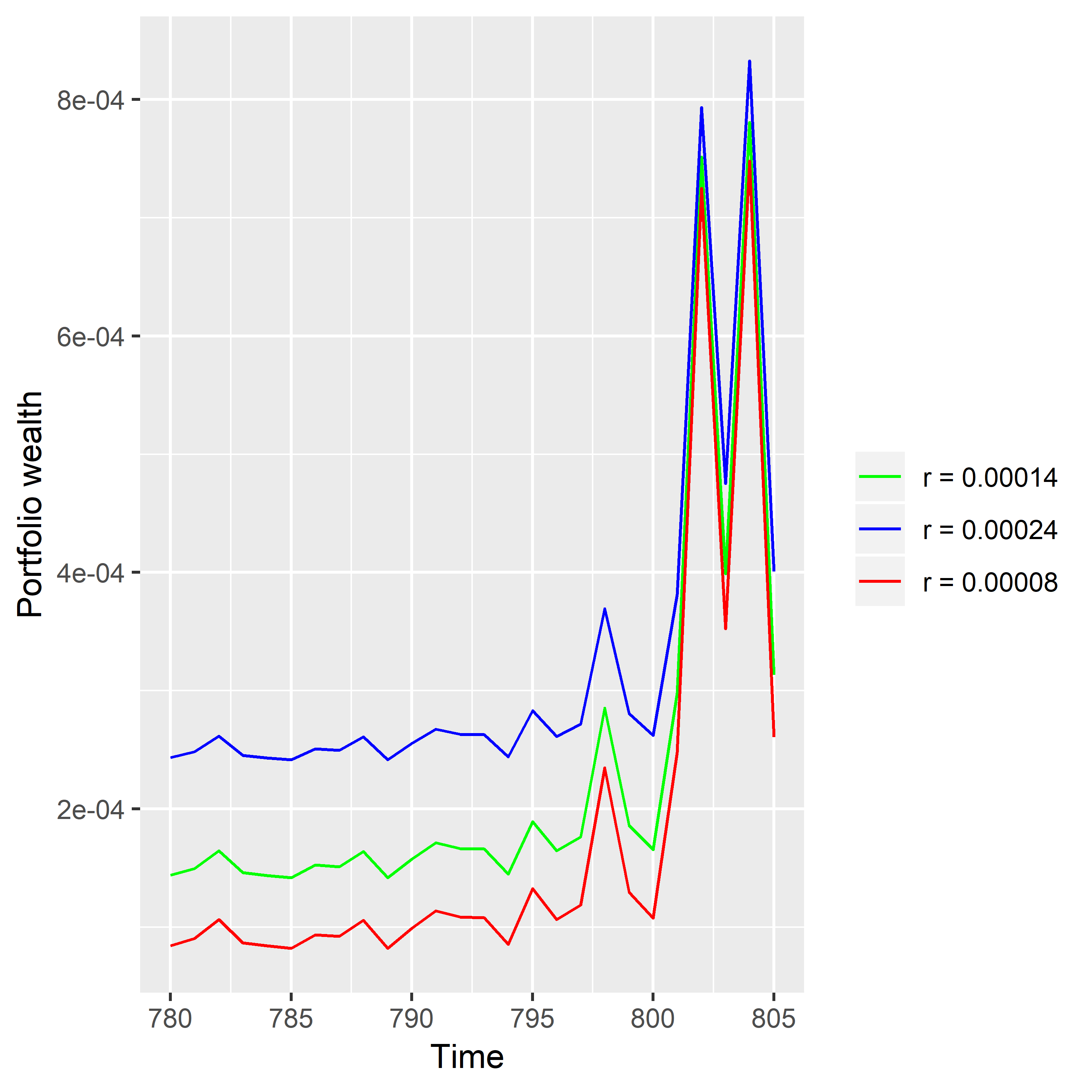}
\caption{When there is no transaction cost }
\label{fig:13}
\end{subfigure}%
\begin{subfigure}{.6\textwidth}
  \centering
  \includegraphics[width=0.95\textwidth]{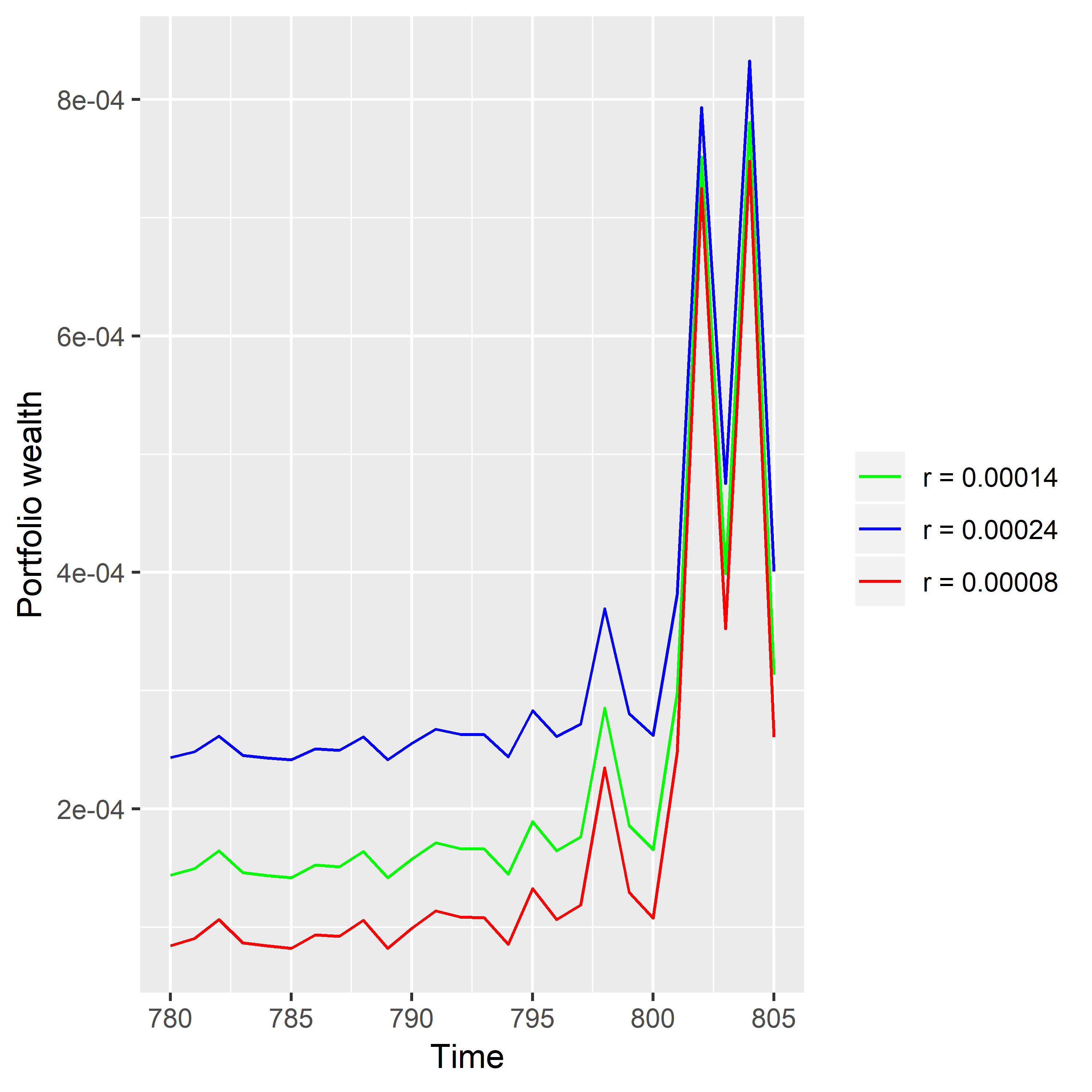}
   \caption{When there is transaction cost }
  \label{fig:13_1}
\end{subfigure}%
\caption{Change in portfolio wealth for different values of r when there is transaction cost and no transaction cost considering Weibull distribution}
\label{fig:test}
\end{figure}

\noindent Finally when we plot the portfolio wealth over the last 26 days of the historical data it can be inferred that throughout the time period, portfolio wealth fluctuates as well as increases for all the values of r, but when the rate of interest r is maximum then the portfolio wealth is also maximum and when the rate of interest is least then the portfolio wealth is minimum (Figure (\ref{fig:13})).\newline
\centerline{\textbf{[Figure (\ref{fig:13})] should be placed here}}

\subsubsection{With transaction cost}

When transaction cost is taken into account the wealth equation \eqref{eq25} modifies to
\begin{equation}
L_t = \pi_{t} S_t + (1 - \pi_{t}) r - (\pi_{t} - \pi_{t-1}) r_1 
\end{equation}
Using \eqref{ing} and taking the value of $r_1$ (Transaction cost rate for a change in optimal strategy) = 10\%, we plot the optimal policy for the last 26 days of the historical data and can observe that the optimal policy increases non-linearly for the entire time period. The change in slope of the curve's are different for different rates of interest. When the rate of interest is least the change in slope is minimum and when the rate of interest is maximum the change in slope is also maximum. (see Figure (\ref{fig:34_1})). The situation is similar to the no transaction cost situation.\newline
\centerline{\textbf{[Figure (\ref{fig:34_1})] should be placed here}} 
\noindent If we plot the value function over the last 26 days of the historical data (Figure (\ref{fig:35_1})) we can perceive that the value function has a similar plot when we compare it to the no transaction cost scenario.\newline
\centerline{\textbf{[Figure (\ref{fig:35_1})] should be placed here}}
\noindent Finally  when  we  plot  the  portfolio  wealth  over the last 26 days of the historical data it can be interpreted that the wealth fluctuates in a similar fashion to that of no transaction cost situation (Figure (\ref{fig:13_1})).\newline
\centerline{\textbf{[Figure (\ref{fig:13_1})] should be placed here}}

\subsection{Comparison of Model Accuracy}

Before we move on to the discussion of our non-parametric analysis, we close this section with a short discussion of a comparison of modelling success between the three candidates we considered and a few other common candidates discussed in literature. We consider the mixture normal distribution (as in \citeauthor{Ref7}, \citeyear{Ref7}) which can easily accommodate fat tails (with finite moments), excess skewness and Kurtosis. We also consider Variance Gamma to consider returns distribution with jumps (as in \citeauthor{Ref25}, \citeyear{Ref25}). We follow the same strategy for testing out of sample goodness of fit as above (once the parameters are estimated using 700 return data points, we check whether the next 50 data points are coming from the same distribution).\newline

\noindent The estimation results for these two distributions are provided below:

\noindent {\bf Mixture Normal:} We consider two components $N(\mu_1, \sigma)$ and $N(\mu_2, \sigma)$. Mixing proportion is $\pi$, standardised distance between the two means is $\delta = |\mu_1 - \mu_2|/\sigma$. Finally, BI is the bimodality index $\delta \sqrt{\pi (1 - \pi)}$ (refer Table \ref{mixnorm}). \newline
\centerline{\textbf{Table \ref{mixnorm} should be here}}

\noindent The KS Test statistic turns out to be $D = 0.083333$ which is less than the critical value of 0.12323. So the distribution fits the data well.\newline
\noindent {\bf Variance Gamma} We use the Nelder- Mead Optimization method to estimate the MLE here (refer Table \ref{vargamm}).\newline
\centerline{\textbf{Table \ref{vargamm} should be here}} 

\noindent Here the KS Test statistic is $D = 0.2603$ which is greater than the critical value. So this candidate distribution fails to fit the out of sample data.

\noindent We summarise the comparison in the table below (refer Table \ref{KS}).\newline
\centerline{\textbf{Table \ref{KS} should be here}} 

\noindent Thus we see that the mixture normal performs the best, although it is much more expensive in that it requires the estimation of six parameters compared to only two for all our candidates. Among them, Inverse Gaussian performs best, followed by Weibull. Variance Gamma fares the worst as it fails to describe the data at hand.

\section{Example and numerical results for unknown distribution of return}

\subsection{Without transaction cost}

When the distribution function is not known for the return of the stock prices we try to fit a distribution using kernel density estimator (KDE). It is a non parametric way to estimate the probability density function of a random variable. KDE is a fundamental data smoothing problem based on the finite data sample we choose. If we have $(x_1, x_2.....x_n)$ as the independent univariate samples coming from an unknown distribution f then we can write 
\begin{equation}
    \widehat{f_h (x)} = \frac{1}{nh}\sum_{i=1}^{n} K \big(\frac{x-x_i}{h}\big) \nonumber
\end{equation}
where K is the kernel which is a non-negative function and h > 0 is a smoothing parameter called the bandwidth \citeauthor{Ref11}(\citeyear{Ref11}). Using the KDE function we get the following estimate of the parameters (refer Table \ref{kernel}).\newline
\centerline{\textbf{Table \ref{kernel} should be here}} 

\begin{figure}
\centering
\begin{subfigure}{.6\textwidth}
  \centering
  \includegraphics[width=0.95\textwidth]{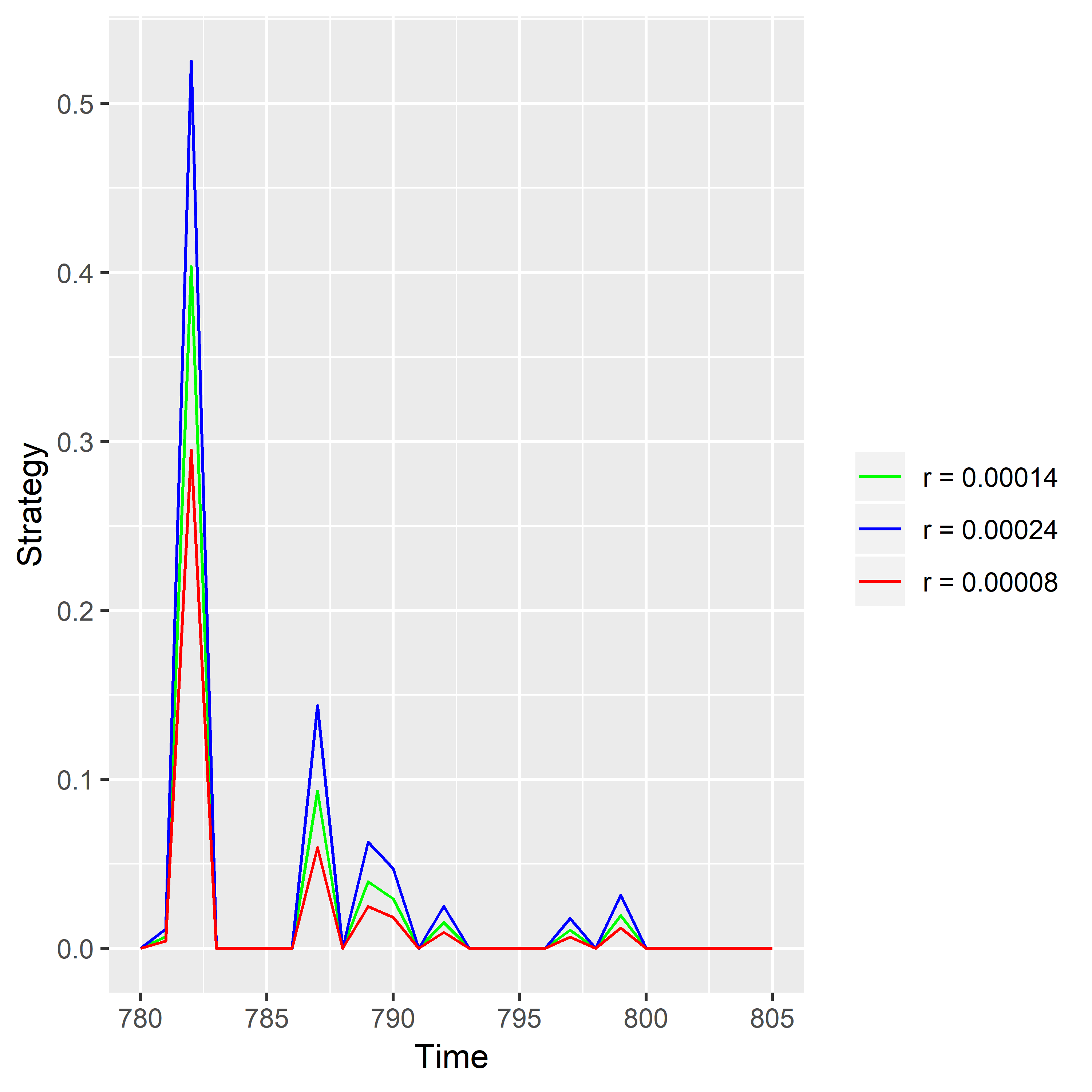}
\caption{When there is no transaction cost }
\label{fig:10}
\end{subfigure}%
\begin{subfigure}{.6\textwidth}
  \centering
  \includegraphics[width=0.95\textwidth]{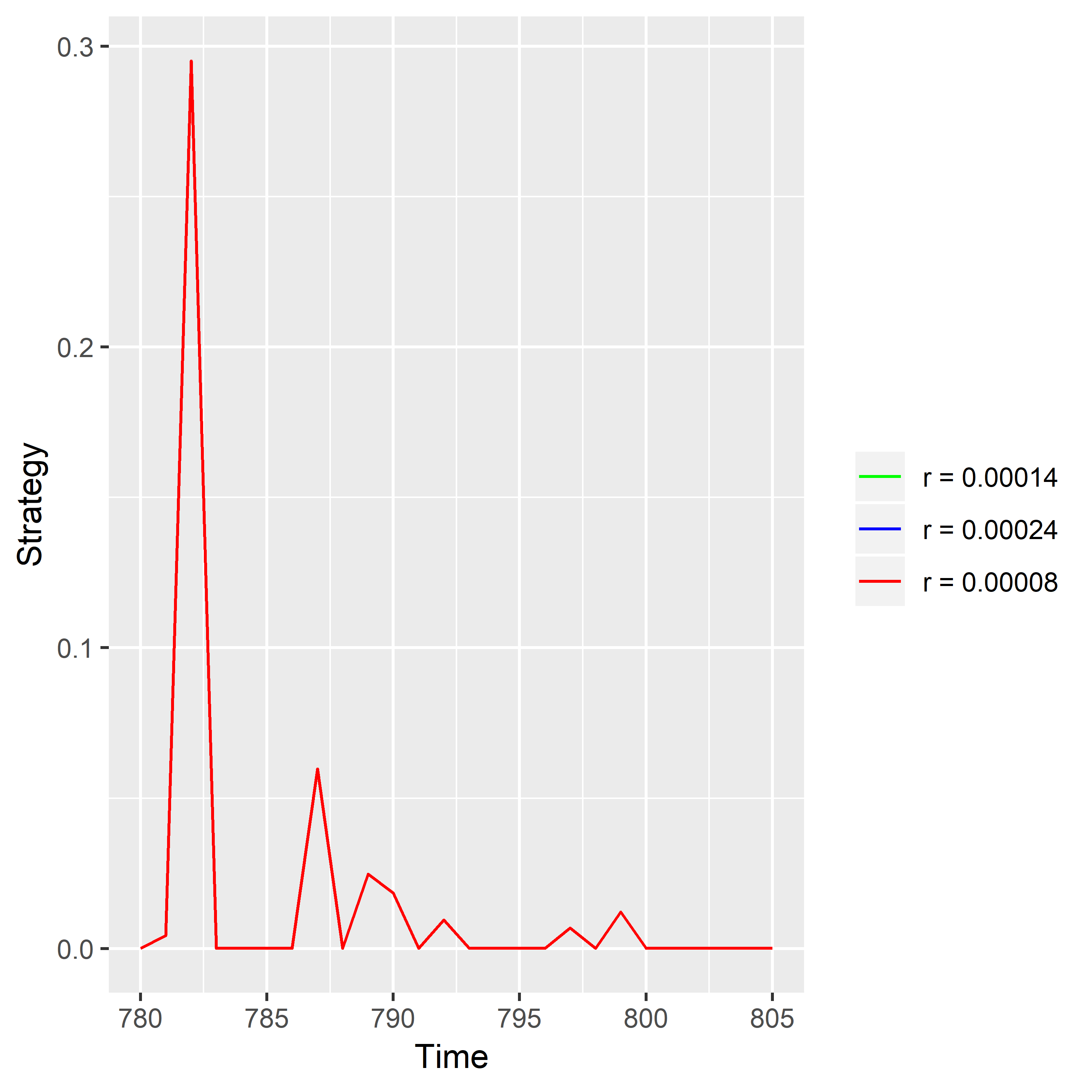}
   \caption{When there is transaction cost }
  \label{fig:10_1}
\end{subfigure}%
\caption{Change in optimal strategy for different values of r when there is transaction cost and no transaction cost considering kernel density estimation}
\label{fig:test}
\end{figure}

\noindent Again we take three different values for r. Now, \eqref{eq29} can be used again for finding the conditional probability of the portfolio wealth. Expanding we get,

\begin{equation}
  \frac{ P (\pi_{t-1}S_{t-1} + (M-\pi_{t-1})r > Q_{0.05}, \pi_{t}S_{t} + (M-\pi_{t})r > Q_{0.05})}{P (\pi_{t}S_{t} + (M-\pi_{t})r > Q_{0.05})} \ge 0.95 \nonumber 
\end{equation}

\begin{equation}
  \frac{ P \big(S_{t-1} > \frac{Q_{0.05} + r(\pi_{t-1}-M)}{\pi_{t-1}}, S_{t} > \frac{Q_{0.05} + r(\pi_{t}-M)}{\pi_{t}}\big)}{P \big(S_{t} > \frac{Q_{0.05} + r(\pi_{t}-M)}{\pi_{t}}\big)} \ge 0.95 \nonumber 
\end{equation}

\begin{equation}\label{eq38}
    \frac{F\big(\frac{Q_{0.05} + r(\pi_{t}-M)}{\pi_{t}}\big)- F\big(\frac{Q_{0.05} + r(\pi_{t-1}-M)}{\pi_{t-1}}\big)}{1 - \big(F\big(\frac{Q_{0.05} + r(\pi_{t-1}-M)}{\pi_{t-1}}\big) - F\big(0\big)\big)} \ge 0.95
\end{equation}
Taking M = 1000 and now solving \eqref{eq38} with the help of standard numerical integration in R-Studio, we get the relationship as depicted in figure (\ref{fig:10}). We can observe that the change in optimal policy is not so frequent for the last 26 days of the historical data. The investor would keep the asset at a particular state for some time. Also for the least rate of interest the peak value of the fluctuation is minimum and for the maximum rate of interest the peak value of the fluctuation is maximum.\newline
\centerline{\textbf{[Figure (\ref{fig:10})] should be placed here}}

\begin{figure}
\centering
\begin{subfigure}{.6\textwidth}
  \centering
  \includegraphics[width=0.95\textwidth]{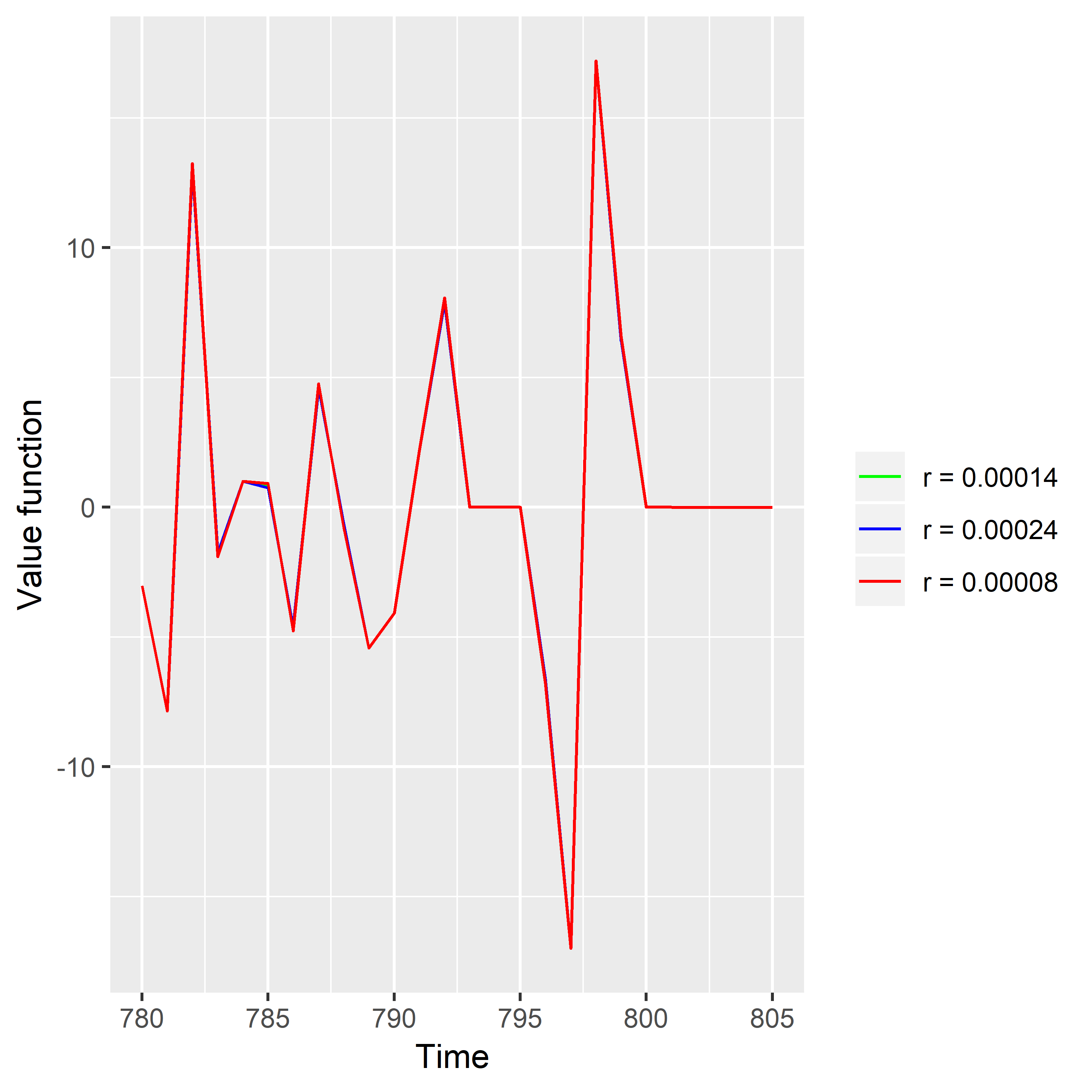}
\caption{When there is no transaction cost }
\label{fig:11}
\end{subfigure}%
\begin{subfigure}{.6\textwidth}
  \centering
  \includegraphics[width=0.95\textwidth]{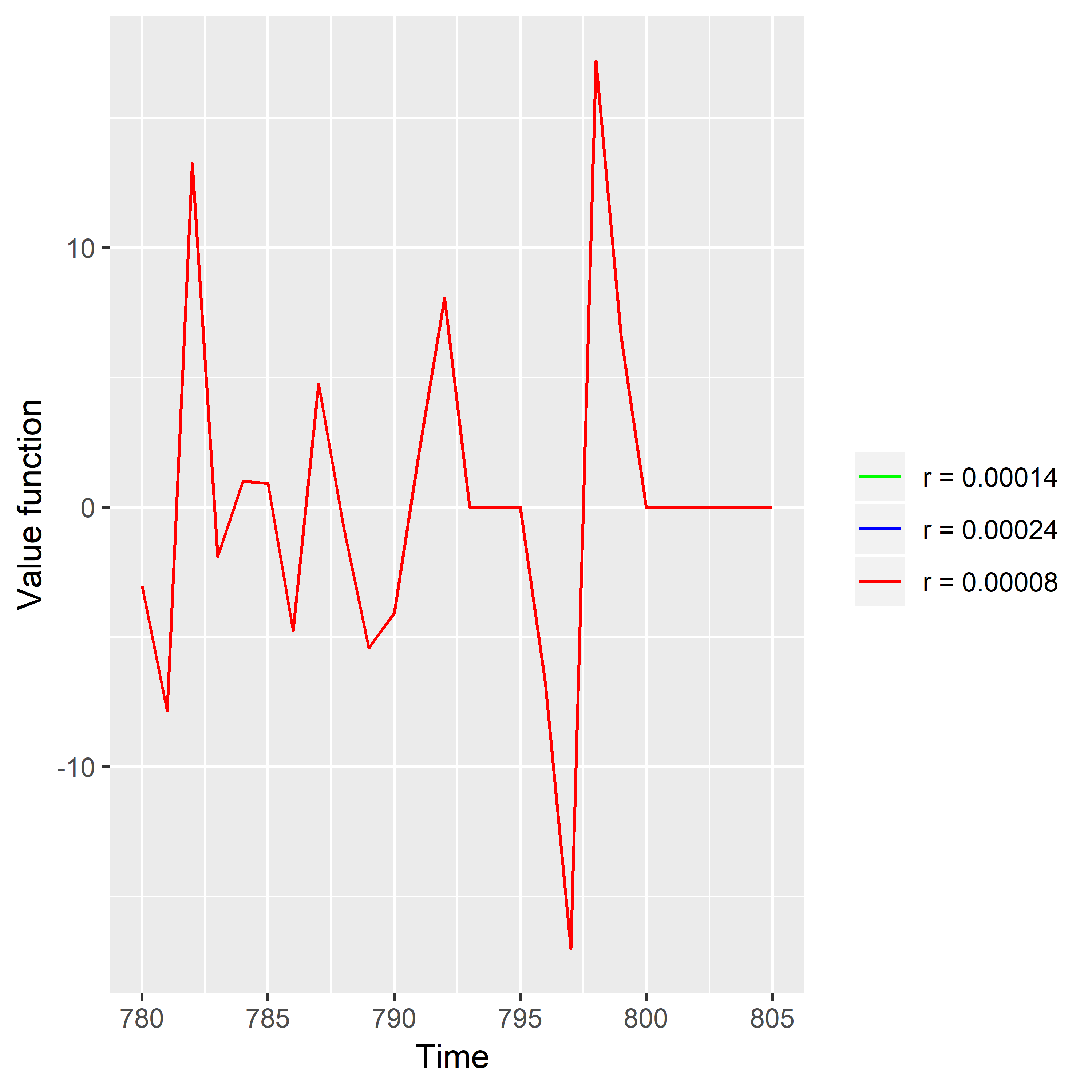}
   \caption{When there is transaction cost }
  \label{fig:11_1}
\end{subfigure}%
\caption{Change in value function for different values of r when there is transaction cost and no transaction cost considering kernel density estimation}
\label{fig:test}
\end{figure}

\noindent If we  consider a terminal value as 0.001 and plot the value function [Figure (\ref{fig:11})], it fluctuates but as it approaches the terminal position it decreases and then increases. Changing the terminal value  doesn't create any difference in the behaviour of the value function. Also the rate of interest has little significance in the plot and the significance can be seen only when it is in the state of fluctuation.\newline
\centerline{\textbf{[Figure (\ref{fig:11})] should be placed here}}

\begin{figure}
\centering
\begin{subfigure}{.6\textwidth}
  \centering
  \includegraphics[width=0.95\textwidth]{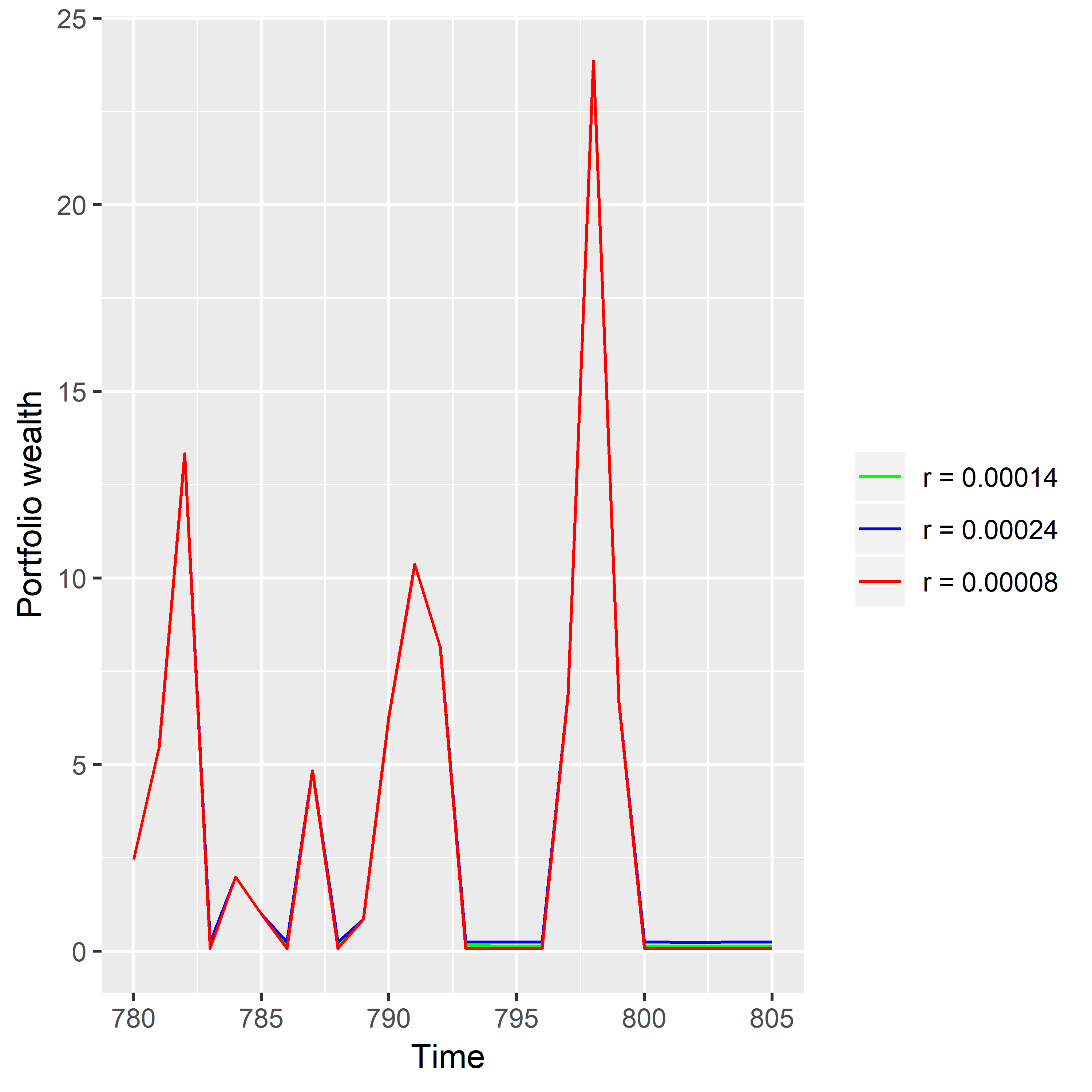}
\caption{When there is no transaction cost }
\label{fig:12}
\end{subfigure}%
\begin{subfigure}{.6\textwidth}
  \centering
  \includegraphics[width=0.95\textwidth]{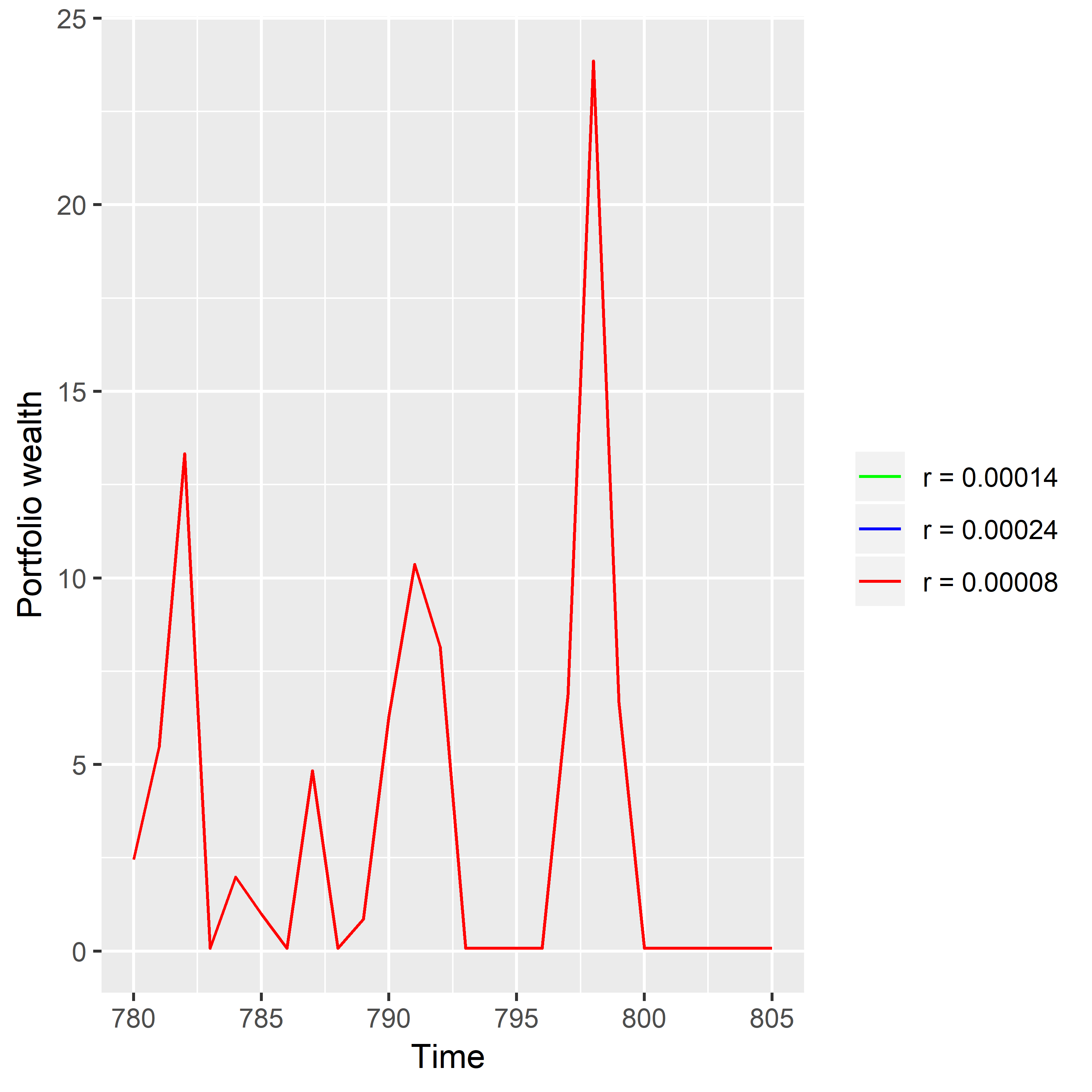}
   \caption{When there is transaction cost }
  \label{fig:12_1}
\end{subfigure}%
\caption{Change in portfolio wealth for different values of r when there is transaction cost and no transaction cost considering  kernel density estimation}
\label{fig:test}
\end{figure}

\noindent Plotting the portfolio wealth [Figure (\ref{fig:12})] we observe that wealth fluctuates but as it approaches the terminal position, the wealth increases in the same manner as the value function. Changing the terminal value doesn't create any difference in the behaviour of the portfolio wealth. Also the change in the rate of interest has very little significance on the plot when the portfolio wealth is constant and during fluctuations it doesn't have any effect.
\centerline{\textbf{[Figure (\ref{fig:12})] should be placed here}}

\subsection{With transaction cost}

Plotting the effect of transaction cost on optimal strategy we can perceive that the fluctuations are similar to the no transaction cost scenario but there is no effect of the interest rate r. (see Figure (\ref{fig:10_1})).\newline
\centerline{\textbf{[Figure (\ref{fig:10_1})] should be placed here}}

\noindent When we are considering transaction cost involved in the process and plot the value function (Figure (\ref{fig:11_1})) we observe that it increases and decreases but as it move towards the terminal position the value function becomes constant. There is no effect of the rate of interest r on the fluctuations.\newline
\centerline{\textbf{[Figure (\ref{fig:11_1})] should be placed here}}

\noindent Similarly when we plot the portfolio wealth (Figure (\ref{fig:12_1})) we observe that the portfolio wealth fluctuates but there is no effect of the interest rate r.\newline
\centerline{\textbf{[Figure (\ref{fig:12_1})] should be placed here}}

\section{Conclusion}

We proposed a new investor objective framework to deal with heavy tail distribution of the return of the stock prices and tried to optimize the portfolio based on managing Value at Risk (Var). As we know that moments do not exist for a heavy tailed distribution, quantiles are the only way to tackle the portfolio optimization problem. In our proposed approach for portfolio optimization we discretized the wealth equation and applied the general Markov Decision Problem formulation. So it has become a dynamic programming problem. We optimize considering two alternative scenarios where the distribution is known and when the distribution is not known. \newline
\noindent In the known distribution case, we have fitted the Pareto, Weibull and Inverse 
Gaussian distributions to the return of the stock prices and found out the quantiles; and when the distribution is not known we used kernel density estimator to find out the quantiles. In both the cases we have shown the recursive relation between the optimal policy and maximized the total expected reward which is the value function at each of the time points. The optimal policy tells us when to build up on the risky asset and when to liquidate it. In the case where we have transaction cost, it will not allow the investor to change its policy very frequently and also not by a large margin. Hence, the value function and portfolio wealth are almost constant but as we approach the terminal position we build up more on the risky asset or liquidate it. 

\noindent In the situation when the density functions are not known, the value function we obtained has wide range of fluctuations and the same is also true for the portfolio wealth. So the investor should come to know when to liquidate on the risky asset and build up on the risky asset. In this case we have used the standard numerical integration method to solve and get the recursive relation for the optimal policy. We have also shown how the transaction cost will affect the policy implementation. Here also the change in the optimal strategy won't be as frequent when transaction cost is present. The results also align with the general intuition when the rate of interest increases or decreases, considering the presence or absence of transaction cost.

\noindent We have also made a comparative discussion of out of sample fit performance among our three candidate distributions and two other distributions commonly considered in the literature, viz. mixture normal and variance gamma. We see that the performance of our candidates are close to the mixture normal with less computational burden (only two parameters to be estimated instead of six). While variance gamma fails to fit the data at all. Thus the distributions functions we considered seem to be quite suitable for empirical work.

\noindent As heavy-tailness is quite common in financial returns data, we hope that our proposed methodology will be useful for practitioners. An additional issue that we have not addressed in this paper is how to incorporate multiple risky assets in this type of analysis. In the absence of moments, the challenge would be to replace the covariance function, which will be undefined, with an appropriate measure of association and carry out the portfolio optimization. We aim to take this issue up in our future work.




\clearpage
\listoftables 
\clearpage

\begin{center}
\begin{tabular}{ | m{3cm} | m{3cm} | m{3cm} | m{3cm} | }
\hline
Parameter &  $\lambda$ & $\alpha$ & $Q_{0.05}$ \\
\hline 
Estimate & 85.34364 & 10346.37374 & $5\times 10^{-6}\lambda$ \\
\hline
Std. Error & 49.85216 & 8537.24358 & -\\
\hline
\end{tabular}
\captionof{table}{Parameter Estimates of Pareto Distribution}\label{pareto}
\end{center}

\begin{center}
\begin{tabular}{| m{3cm} | m{3cm} | m{3cm} | m{3cm} | }
\hline
Parameters & $\lambda$ & $\alpha$ & $Q_{0.05}$ \\
\hline
Estimate & 1.2630 & 0.0104 & $0.00078\lambda$ \\
\hline
Std. Error & 0.06463 & 0.00054 & - \\
\hline
\end{tabular}
\captionof{table}{Parameter Estimates of Weibull Distribution}\label{weibull}
\end{center}

\begin{center}
\begin{tabular}{| m{3cm} | m{3cm} | m{3cm} | m{3cm} | }
\hline 
Parameters & $mean(\mu)$ & $ shape(\lambda)$ & $Q_{0.05}$ \\
\hline 
Estimate & 0.0097 & 0.0044 & 0.00096 \\
\hline
Std. Error & 0.00087 & 0.00038 & - \\
\hline
\end{tabular}
\captionof{table}{Parameter Estimates of Inverse Gaussian Distribution}\label{invgauss}
\end{center}

\begin{center}
\begin{tabular}{|c|c|c|c|c|c|c|}
\hline
Parameter & $\mu_1$ & $\mu_2$ & $\sigma$ & $\delta$ & $\pi$ & BI \\
\hline
Estimate & 0.007286 & 0.02137 & 0.004741 & 2.97 & 0.8646 & 1.016 \\
\hline 
\end{tabular} 
\captionof{table}{Parameter Estimates of Mixture Normal Distribution}\label{mixnorm}
\end{center}

\begin{center}
\begin{tabular}{|c|c|c|c|c|}
\hline
Parameter & $c$ (location) & $\sigma$ (spread) & $\theta$ (asymmetry) & $\nu$ (shape)\\
\hline
Estimate & 0.005034 & 0.017160 & 0.006051 & 3.219203 \\
\hline
\end{tabular}
\captionof{table}{Parameter Estimates of Variance Gamma Distribution}\label{vargamm}
\end{center}

\begin{center}
\begin{tabular}{|c|c|c|c|c|c|}
\hline
Distribution & Pareto & Weibull & Inverse Gaussian & Mixture Normal (2 components) & Variance Gamma \\
\hline
Number of Parameters  & 2 & 2 & 2 & 6 & 4 \\
\hline
$D$ statistic & 0.16816 & 0.152271 & 0.14623 & 0.083333 & 0.2603 \\
\hline
critical value & 0.19206 & 0.19206 & 0.19206 & 0.12323 & 0.156 \\
\hline
\end{tabular}
\captionof{table}{KS comparison test of all the distribution}\label{KS}
\end{center}

\begin{center}
\begin{tabular}{| m{5cm} | m{3cm} | }
\hline
kernel density estimate & log-quadratic fitting \\
\hline
bandwidth (bw) & 0.00271447 \\
\hline
\end{tabular}
\captionof{table}{Parameter Estimates of Kernel density estimator}\label{kernel}
\end{center}

\end{document}